\documentclass[12pt]{article}

\usepackage[utf8]{inputenc}
\usepackage{charter}
\usepackage{fullpage}
\usepackage{graphicx}
\usepackage{lipsum}
\usepackage[sort&compress,numbers]{natbib}
\usepackage[total={6.5in,9in},top=1in,headsep=0.1in,headheight=1in]{geometry}

\usepackage{times}
\usepackage{enumitem,amssymb}
\usepackage{ragged2e}
\usepackage{graphicx}
\usepackage{comment}
\usepackage[usenames]{xcolor} 
\usepackage[vietnamese,english]{babel}
\usepackage[outercaption]{sidecap}
\usepackage{tabularx}
\usepackage{multirow}
\usepackage{threeparttable}
\usepackage{setspace} 
\usepackage{caption} 
\renewcommand{\arraystretch}{1.3} 

\definecolor{xlinkcolor}{cmyk}{1,1,0,0}
\usepackage{url}
\usepackage[
 colorlinks=true,    
 linkcolor=xlinkcolor,     
 citecolor=xlinkcolor,     
 filecolor=xlinkcolor,  
 urlcolor=xlinkcolor,      
 final=true
]{hyperref}

\hypersetup{hidelinks}

\newcommand{\wde}{w_{0}}
\newcommand{\summnu}{\sum m_{\nu}}
\newcommand{\mvir}{M_{500c}}
\newcommand{\msol}{{\rm M}_{\odot}}
\newcommand{\taure}{\tau_{\rm re}}

\newcommand\snowmass{
\begin{center}
  \rule[-0.2in]{\hsize}{0.01in}\\
  \rule{\hsize}{0.01in}\\
  \vskip 0.1in
  Submitted to the Proceedings of the US Community Study\\ 
  on the Future of Particle Physics (Snowmass 2021)\\
  \rule{\hsize}{0.01in}\\
  \rule[+0.2in]{\hsize}{0.01in}\\[-2em]
\end{center}
}

\usepackage[firstpage=true]{background}
\backgroundsetup{contents={\parbox{6.5in}{\snowmass}}, scale=1,placement=top,opacity=1,color=black,position={3.25in,1.2in}}

\usepackage{fancyhdr}
\fancypagestyle{plain}{%
  \fancyhf{}%
  \fancyhead[C]{}
  \fancyfoot[C]{\thepage}
}

\fancypagestyle{empty}{%
  \fancyhf{}%
  \fancyhead[C]{{\it Snowmass 2021 CMB-HD White Paper}}
  \fancyfoot[C]{\thepage}
}
\definecolor{ao(english)}{rgb}{0.0, 0.5, 0.0}

\usepackage{sectsty}

\definecolor{DarkGreen}{rgb}{0.0, 0.3, 0.0}
\definecolor{purple}{rgb}{0.5, 0.0, 0.5}
\definecolor{red}{rgb}{1, 0.0, 0.0}
\definecolor{green}{rgb}{0, 1.0, 0.0}

\newcommand{\apjl}{\rm ApJL}
\newcommand{\apjs}{\rm ApJS}

\newcommand{\aap}{\rm A$\&$A}

\def\ao{{\it Appl.~Opt.}}           

\def\3he{$^3{\rm He}$}

\def\lsim{\mathrel{\lower2.5pt\vbox{\lineskip=0pt\baselineskip=0pt
           \hbox{$<$}\hbox{$\sim$}}}}

\def\gsim{\mathrel{\lower2.5pt\vbox{\lineskip=0pt\baselineskip=0pt
           \hbox{$>$}\hbox{$\sim$}}}}

\title{Snowmass2021 CMB-HD White Paper}
\date{}

\usepackage{authblk}

\begin{document}

\author[1]{The CMB-HD Collaboration}
\author[2]{Simone Aiola}
\author[3,4,5,6]{Yashar Akrami}
\author[7]{Kaustuv Basu}
\author[8]{Michael Boylan-Kolchin}
\author[9,10]{Thejs Brinckmann}
\author[11]{Sean Bryan}
\author[8]{Caitlin M Casey}
\author[12]{Jens Chluba}
\author[13]{Sébastien Clesse}
\author[14]{Francis-Yan Cyr-Racine}
\author[15,16,17]{Luca Di Mascolo}
\author[18]{Simon Dicker}
\author[19]{Thomas Essinger-Hileman}
\author[20,21]{Gerrit S. Farren}
\author[22]{Michael A.~Fedderke}
\author[23, 24]{Simone Ferraro}
\author[25]{George M. Fuller}
\author[26]{Nicholas Galitzki} 
\author[27]{Vera Gluscevic}
\author[21]{Daniel Grin}
\author[20]{Dongwon Han}
\author[2]{Matthew Hasselfield}
\author[28,29]{Ren\'ee Hlo\v{z}ek}
\author[30]{Gil Holder}
\author[22]{Selim~C.~Hotinli}
\author[18]{Bhuvnesh Jain}
\author[31]{Bradley Johnson}
\author[32,33]{Matthew Johnson}
\author[34]{Pamela Klaassen}
\author[35]{Amanda MacInnis}
\author[33,27]{Mathew Madhavacheril}
\author[35]{Sayan Mandal}
\author[36,37]{Philip Mauskopf}
\author[38]{Daan Meerburg}
\author[39]{Joel Meyers}
\author[35]{Vivian Miranda}
\author[40]{Tony Mroczkowski}
\author[33,41,42,43]{Suvodip Mukherjee}
\author[44]{Moritz M{\"u}nchmeyer}
\author[45]{Julian Munoz}
\author[2]{Sigurd Naess}
\author[46]{Daisuke Nagai}
\author[47]{Toshiya Namikawa}
\author[46]{Laura Newburgh}
\author[48]{\foreignlanguage{vietnamese}{Hồ Nam Nguyễn}}
\author[49]{Michael Niemack}
\author[50,45]{Benjamin D. Oppenheimer}
\author[27]{Elena Pierpaoli}
\author[51]{Srinivasan Raghunathan}
\author[23,24]{Emmanuel Schaan}
\author[35]{Neelima Sehgal}
\author[20]{Blake Sherwin}
\author[52]{Sara M. Simon}
\author[53]{An\v{z}e Slosar}
\author[33]{Kendrick Smith}
\author[2]{David Spergel}
\author[19]{Eric R.~Switzer}
\author[54]{Pranjal Trivedi}
\author[55]{Yu-Dai Tsai}
\author[11]{Alexander van Engelen}
\author[2,56,57]{Benjamin D.~Wandelt}
\author[19]{Edward J. Wollack}
\author[58,59]{Kimmy Wu}

\affil[1]{http://cmb-hd.org} 
\affil[2]{Center for Computational Astrophysics, Flatiron Institute, 162 5th Avenue, New York, NY 10010 USA} 
\affil[3]{CERCA/ISO, Department of Physics, Case Western Reserve University, 10900 Euclid Avenue, Cleveland, OH 44106, USA}
\affil[4]{Department of Physics, Imperial College London, Blackett Laboratory,
Prince Consort Road, London SW7 2AZ, United Kingdom}
\affil[5]{Laboratoire de Physique de l'\'Ecole Normale Sup\'erieure, ENS, Universit\'e PSL, CNRS, Sorbonne Universit\'e, 
F-75005 Paris, France}
\affil[6]{Observatoire de Paris, Universit\'e PSL, Sorbonne Universit\'e, LERMA, 750 Paris, France}
\affil[7]{Argelander Institute for Astronomy, University of Bonn, Auf dem Hügel 71, D-53121 Bonn, Germany}
\affil[8]{Department of Astronomy, The University of Texas at Austin, 2515 Speedway, Stop C1400, Austin, Texas 78712, USA}
\affil[9]{Dipartimento di Fisica e Scienze della Terra, Universit\'a degli Studi di Ferrara, via Giuseppe Saragat 1, 44122 Ferrara, Italy}
\affil[10]{Istituto Nazionale di Fisica Nucleare (INFN), Sezione di Ferrara, Via Giuseppe Saragat 1, 44122 Ferrara, Italy}
\affil[11]{School of Earth and Space Exploration, Arizona State University, 781 Terrace Mall, Tempe, AZ 85287, USA}
\affil[12]{Jodrell Bank Centre for Astrophysics, Alan Turing Building, University of Manchester, Manchester M13 9PL, United Kingdom}
\affil[13]{Service de Physique Théorique (CP225), Université Libre de Bruxelles, Boulevard du Triomphe, 1050 Brussels, Belgium}
\affil[14]{Department of Physics and Astronomy, University of New Mexico, 210 Yale Blvd NE, Albuquerque, NM 87106, USA}
\affil[15]{Astronomy Unit, Department of Physics, University of Trieste, via Tiepolo 11, Trieste 34131, Italy}
\affil[16]{INAF - Osservatorio Astronomico di Trieste, via Tiepolo 11, Trieste 34131, Italy}
\affil[17]{IFPU - Institute for Fundamental Physics of the Universe, Via Beirut 2, 340 Trieste, Italy}
\affil[18]{Department of Physics and Astronomy, University of Pennsylvania, 209 South 33rd Street, Philadelphia, PA 19104, USA}
\affil[19]{NASA/Goddard Space Flight Center, Greenbelt, MD 20771, USA}
\affil[20]{Department of Applied Mathematics and Theoretical Physics, University of Cambridge, Cambridge CB3 0WA, United Kingdom}
\affil[21]{Department of Physics and Astronomy, Haverford College,
370 Lancaster Avenue, Haverford, Pennsylvania 19041, USA}
\affil[22]{The William H.~Miller III Department of Physics and Astronomy, The Johns Hopkins University, Baltimore, MD 21218, USA}
\affil[23]{Lawrence Berkeley National Laboratory, One Cyclotron Road, Berkeley, CA 94720, USA}
\affil[24]{Berkeley Center for Cosmological Physics, Department of Physics,
University of California, Berkeley, CA 94720, USA}
\affil[25]{Center for Astrophysics and Space Sciences, University of California, San Diego, La Jolla, CA 92093, USA}
\affil[26]{University of California, San Diego, Department of Physics, San Diego, CA 92093, USA}
\affil[27]{Department of Physics and Astronomy, University of Southern California, Los Angeles, CA, 90007, USA}
\affil[28]{Dunlap Institute for Astronomy and Astrophysics, University of Toronto,
50 St George Street, Toronto, Ontario, M5S 3H4, Canada}
\affil[29]{David A. Dunlap Department of Astronomy and Astrophysics,
University of Toronto, 50 St George Street, Toronto, Ontario, M5S 3H4, Canada}
\affil[30]{University of Illinois at Urbana-Champaign, 1110 W Green, Urbana IL 61801, USA}
\affil[31]{Department of Astronomy, University of Virginia, Charlottesville, VA 22904, USA}
\affil[32]{Department of Physics and Astronomy, York University, Toronto, ON M3J 1P3, Canada}
\affil[33]{Perimeter Institute for Theoretical Physics, 31 Caroline St N, Waterloo, ON N2L 2Y5, Canada}
\affil[34]{UK Astronomy Technology Centre, Blackford Hill, Edinburgh EH9 3HJ, United Kingdom}
\affil[35]{Physics and Astronomy Department, Stony Brook University, Stony Brook, NY 11794, USA}
\affil[36]{School of Physics and Astronomy, Cardiff University, Queen’s Buildings, The Parade, Cardiff, CF24 3AA, United Kingdom}
\affil[37]{School of Earth and Space Exploration and Department of Physics, Arizona State University, Tempe, AZ 85287, USA}
\affil[38]{Van Swinderen Institute for Particle Physics and Gravity, University of Groningen, Nijen- borgh 4, 9747 AG Groningen, The Netherlands}
\affil[39]{Department of Physics, Southern Methodist University, 3215 Daniel Ave, Dallas, TX 75275, USA}
\affil[40]{European Southern Observatory (ESO), Karl- Schwarzschild-Strasse 2, Garching 85748, Germany}
\affil[41]{Gravitation Astroparticle Physics Amsterdam (GRAPPA), Anton Pannekoek Institute for Astronomy and Institute for Physics, University of Amsterdam, Science Park 904, 1090 GL Amsterdam, The Netherlands}
\affil[42]{Institute Lorentz, Leiden University, PO Box 9506, Leiden 2300 RA, The Netherlands}
\affil[43]{Delta Institute for Theoretical Physics, Science Park 904, 1090 GL Amsterdam, The Netherlands}
\affil[44]{Department of Physics, University of Wisconsin-Madison, Madison, WI 53706, USA}
\affil[45]{Harvard-Smithsonian Center for Astrophysics, 60 Garden St., Cambridge, MA 02138, USA}
\affil[46]{Department of Physics, Yale University, P.O. Box 208120, New Haven, CT, 06520, USA}
\affil[47]{Kavli Institute for the Physics and Mathematics of the Universe (WPI), UTIAS, The University of Tokyo, Kashiwa, Chiba 277-8583, Japan}
\affil[48]{Department of Physics, University of California, Berkeley, CA 94720, USA}
\affil[49]{Cornell University Physics Department and Astronomy Department, Ithaca, NY 853 USA}
\affil[50]{CASA, Department of Astrophysical and Planetary Sciences, University of Colorado, 389 UCB, Boulder, CO 80309, USA}
\affil[51]{Center for AstroPhysical Surveys, National Center for Supercomputing Applications, Urbana, IL 61801, USA}
\affil[52]{Fermi National Accelerator Laboratory, MS209, P.O. Box 500, Batavia, IL, 60510, USA}
\affil[53]{Physics Department, Brookhaven National Laboratory, Upton, NY 11973, USA}
\affil[54]{Hamburger Sternwarte, University of Hamburg, Gojenbergsweg 112, D-21029 Hamburg, Germany}
\affil[55]{Department of Physics and Astronomy, University of California, Irvine, CA 92697-4575, USA}
\affil[56]{Sorbonne Université, CNRS, UMR 7095, Institut d’Astrophysique de Paris (IAP), 98 bis bd Arago, F-750 Paris, France}
\affil[57]{Sorbonne Université, Institut Lagrange de Paris (ILP), 98 bis bd Arago, F-750 Paris, France}
\affil[58]{Kavli Institute for Particle Astrophysics and Cosmology, Stanford University, 452 Lomita Mall, Stanford, CA 94305, USA}
\affil[59]{SLAC National Accelerator Laboratory, 2575 Sand Hill Road, Menlo Park, CA 94025, USA}

\maketitle
\pagenumbering{gobble}
\pagenumbering{arabic}
\thispagestyle{empty}
\setcounter{page}{1}

\vspace{10mm}
\begin{abstract}
\noindent
CMB-HD is a proposed millimeter-wave survey over half the sky that would be ultra-deep (0.5 $\mu$K-arcmin) and have unprecedented resolution (15 arcseconds at 150 GHz).  Such a survey would answer many outstanding questions about the fundamental physics of the Universe. Major advances would be 1.)~the use of gravitational lensing of the primordial microwave background to map the distribution of matter on small scales ($k\sim10~h$Mpc$^{-1}$), which probes dark matter particle properties.  It will also allow 2.)~measurements of the thermal and kinetic Sunyaev-Zel'dovich effects on small scales to map the gas density and velocity, another probe of cosmic structure.  In addition, CMB-HD would allow us to cross critical thresholds: 3.)~ruling out or detecting any new, light ($< 0.1$ eV) particles that were in thermal equilibrium with known particles in the early Universe, 4.)~testing a wide class of multi-field models that could explain an epoch of inflation in the early Universe, and 5.)~ruling out or detecting inflationary magnetic fields.  CMB-HD would also provide world-leading constraints on 6.)~axion-like particles, 7.)~cosmic birefringence, 8.)~the sum of the neutrino masses, and 9.)~the dark energy equation of state.  The CMB-HD survey would be delivered in 7.5 years of observing 20,000 square degrees of sky, using two new 30-meter-class off-axis crossed Dragone telescopes to be located at Cerro Toco in the Atacama Desert. Each telescope would field 800,000 detectors (200,000 pixels), for a total of 1.6 million detectors. 
\end{abstract}

\clearpage
\section{Executive Summary}

CMB-HD is an ambitious leap beyond previous and upcoming ground-based millimeter-wave experiments. It will allow us to cross critical measurement thresholds and definitively answer pressing questions in both astrophysics and the fundamental physics of the Universe.  The combination of CMB-HD with contemporary ground and space-based experiments will also provide countless powerful synergies.

Two critical advances uniquely enabled by CMB-HD are mapping over half the sky:~i)~the distribution of all matter on small scales ($k\sim10~h$Mpc$^{-1}$) using the gravitational lensing of the cosmic microwave background (CMB), and~ii)~the distribution of gas density and velocity on small scales using the thermal and kinetic Sunyaev-Zel'dovich effects (tSZ and kSZ).  The combination of high-resolution and seven frequency bands in the range of 30 to 350 GHz allows for separation of foregrounds from the CMB.  That plus the unprecedented depth of the survey over half the sky allows CMB-HD to achieve the key science targets summarized in Table~\ref{tab:goals}.  This white paper outlines the key science goals motivating the CMB-HD survey, as well as the flowdown to measurement and instrument requirements.  Additional details are discussed in the Astro2020 Science White Paper~\cite{Sehgal2019a}, Astro2020 CMB-HD APC~\cite{Sehgal2019b}, and Astro2020 CMB-HD RFI~\cite{Sehgal2019c}.  Further information can also be found at~\href{https://cmb-hd.org}{https://cmb-hd.org}.

\clearpage
\pagenumbering{gobble}
\linespread{1}
\begin{table*}[h]
\centering
\caption[Key Science]{Summary of CMB-HD Key Science Goals} 
\vspace{0.4mm}
\begin{tabular}{l l l}
\hline
\hline
Science & \hspace{-0.5cm} Parameter &  Sensitivity\\
\hline 
\vspace{1.5mm}
{\bf{Dark Matter \& Dark Sectors}} &   & \\ 
\hspace{0.4cm} Warm Dark Matter & $m_{\rm{WDM}}$  & $m_{\rm{WDM}} < 1$ keV with S/N $> 5\sigma$ \\
\hspace{0.4cm} Fuzzy Dark Matter & $m_{\rm{FDM}}$  & $m_{\rm{FDM}} < 1\times 10^{-22}$ eV with S/N $> 5\sigma$ \\
\vspace{2mm}
\hspace{0.4cm} Axion-Like Particles\textsuperscript{a} &  $g_{ a \gamma}$ & $g_{ a \gamma}<2.1\times 10^{-15}$~GeV$^{-1} (m_a/10^{-22}~\rm{eV})$ (95\% CL)\\
& & $g_{ a \gamma}< 10^{-13}$~GeV$^{-1}$ (95\% CL)\\
\vspace{2mm}
& & for $10^{-13}$~eV~$<m_a\lesssim 2\times 10^{-12}$~eV\\
\hline
\vspace{1mm}

{\bf{Light Relic Particles}}  &   & \\ 
\vspace{1.5mm}
\hspace{0.4cm} Free-streaming Relativistic & $N_{\rm{eff}}$ & $\sigma(N_{\rm{eff}}) = 0.014$ \\
\hline
\vspace{1mm}

{\bf{Inflation}}  &   & \\
\hspace{0.4cm} Inflationary Magnetic Fields & $B_{\rm{SI}}$  & $\sigma(B_{\rm{SI}}) = 0.036$ nG  \\ 
\hspace{0.4cm} Primordial Non-Gaussianity & $f_\mathrm{NL}^{\rm local}$  & $\sigma(f_\mathrm{NL}^{\rm local}) = 0.26$ \\ 
\vspace{2mm}
\hspace{0.4cm} Primordial Gravitational Waves & $r$  & $\sigma(r) = 0.005$ \\ 
\hline
\vspace{1mm}

{\bf{Dark Energy}} &	 &  \\
\vspace{1.5mm}
\hspace{0.4cm} Dark Energy Equation of State & $w$  & $\sigma({w})=0.005$  \\ 
\hline
\vspace{1mm}

{\bf{Neutrino Mass}} &	 &  \\
\vspace{1.5mm}
\hspace{0.4cm} Sum of Neutrino Masses & $\Sigma m_\nu$  & $\sigma({\Sigma m_\nu})=13$ meV \\ 
\hline
\vspace{1mm}

{\bf{Other Beyond Standard Model}} &	 &  \\
\hspace{0.4cm} Isotropic Birefringence & $\beta$  & $\sigma({\beta})=0.035$ arcmin \\ 
\vspace{2mm}
\hspace{0.4cm} Anisotropic Birefringence& $A_{\alpha}$ & $\sigma(A_{\alpha}) = 1.4 \times 10^{-6}$ deg$^2$ \\
\hline
\vspace{1mm}

\vspace{1.5mm}
{\bf{Astrophysics}} & &  \\
\hspace{0.4cm} Planetary Studies & &  Detect dwarf-size planets 100s of AU from Sun\\
& & Detect Earth-sized planets 1000s of AU from Sun\\
\vspace{2mm}
& & Detect exo-Oort clouds around other stars\\
\vspace{-1mm}
\hspace{0.4cm} Evolution of Gas & &  Probe gas physics of halos out to $z\sim2$\\ 
\vspace{2mm}
& & and with masses below $10^{12}$ M$_{\odot}$\\
\vspace{2mm}
\hspace{0.4cm} Transient Sky & & Source flux~$ > 4$ mJy at 150 GHz ($8\sigma$ daily limit) \\
\vspace{2mm}
\hspace{0.4cm} Catalog of Galaxies &  & Source flux~$ > 0.5$ mJy at 150 GHz ($10\sigma$ limit) \\
\hline
\hline
\end{tabular}
\begin{tablenotes}
\item \textsuperscript{a} CMB-HD has several methods to constrain axion-like particles (see Sections~\ref{sec:grav},\ref{sec:axion-like1}, and \ref{sec:axion-like2}). 
\end{tablenotes}
\label{tab:goals}
\vspace{-3mm}
\end{table*}

\clearpage
\pagenumbering{arabic}
\setcounter{page}{3}
\setcounter{tocdepth}{2} 
\tableofcontents

\clearpage

\section{Overview of CMB-HD Key Science Objectives}

\subsection{Fundamental Physics of the Universe} 

\vspace{2mm}
\noindent CMB-HD has the following science objectives in regard to fundamental physics of the Universe:
\vspace{2mm}

\begin{enumerate}
\item {\bf{Dark Matter:}} Measure the small-scale matter power spectrum from weak gravitational lensing using the CMB as a backlight; with this CMB-HD aims to distinguish between a matter power spectrum predicted by models that can explain observational puzzles of small-scale structure, and that predicted by vanilla cold dark matter (CDM), with a significance of at least $5\sigma$~\cite{Han:2021vtm}. This measurement would be a clean measurement of the matter power spectrum on these scales, free of the use of baryonic tracers. It would greatly limit the allowed models of dark matter and baryonic physics, shedding light on dark-matter particle properties and galaxy evolution~\cite{Hlozek:2016lzm,Hlozek:2017zzf,Nguyen2019}. Specifically, this measurement would constrain fuzzy dark matter, warm dark matter, self-interacting dark matter, and any other dark matter model that alters the matter power spectrum on scales of $k\sim10~$Mpc$^{-1}$~\cite{Markovic:2013iza,Hlozek:2016lzm,Hlozek:2017zzf,Li:2018zdm,Nguyen2019}.

Another observable emerges from the epoch of re-ionization. Compton up-scattering of incident CMB photons by plasma along the line of sight is the well known kinetic Sunyaev-Zeldovich (kSZ) effect. When induced by mildly non-linear structure, this Ostriker-Vishniac (kSZ-OV) effect (which directly probes the late-time growth of structure) leads to additional power in the CMB damping tail. By suppressing the formation structure, alternate dark matter models would alter the shape of kSZ-OV-induced anisotropies \cite{Farren:2021jcd}, potentially facilitating a $2.7\times 10^{-21}~{\rm eV}$ 95\% CL lower limit to the fuzzy dark matter (FDM) mass.\\

\item {\bf{Axion-like Particles:}} Constrain or discover axion-like particles by observing the resonant conversion of CMB photons into axions-like particles in the magnetic fields of galaxy clusters. Nearly massless pseudoscalar bosons, often generically called axion-like particles, appear in many extensions of the standard model~\cite{PhysRevLett.38.1440,PhysRevLett.40.223,PhysRevLett.40.279,Svrcek:2006yi,Arvanitaki:2009fg,Acharya:2010zx}. A detection would have major implications both for particle physics and for cosmology, not least because axion-like particles are also a well-motivated dark matter candidate. CMB-HD has the opportunity to provide a world-leading  probe of the electromagnetic interaction between axions-like particles and photons using the resonant conversion of CMB photons into axions-like particles~\cite{Raffelt:1996wa,Mukherjee:2018oeb, Mukherjee:2019dsu} in the magnetic field of galaxy clusters~\cite{Mukherjee:2019dsu}, independently of whether axion-like particles constitute the dark matter.  CMB-HD would explore the mass range of $10^{-13}$~eV~$<m_a\lesssim 2\times 10^{-12}$~eV and improve the constraint on the axion coupling constant by over two orders of magnitude over current particle physics constraints to $g_{ a \gamma}<0.1\times 10^{-12}$~GeV$^{-1}$. These ranges are unexplored to date and complementary with other cosmological searches for the imprints of axion-like particles on the cosmic density field. \\

\item {\bf{Axion-like Particles:}} Constrain or discover axion-like dark matter by measuring time-dependent CMB polarization rotation. Ultralight axion-like dark-matter fields that couple to photons via $g_{a\gamma}$ cause a time-dependent photon birefringence effect which manifests as a temporal oscillation of the local CMB polarization angle (i.e., a local $Q\leftrightarrow U$ oscillation in time)~\cite{Fedderke:2019ajk}.
This rotation effect is in-phase across the entire sky, and the oscillation period is fixed by the axion-like particle mass (a fundamental physics parameter) to be at observable timescales of $\sim$months to $\sim$hours for masses in the range $10^{-21}\,\textrm{eV}\lesssim m_a \lesssim 10^{-18}\,\textrm{eV}$. Being a time-dependent oscillation of the \emph{observed} CMB polarization pattern, searches for this effect are not limited by cosmic variance. CMB-HD improvements in the polarization map-depth and sky coverage promise sensitivity improvements for this search by a couple of orders of magnitude in the coupling $g_{a\gamma}$ as compared to existing BICEP/Keck analyses~\cite{BICEPKeck:2020hhe,BICEPKeck:2021sbt}, and exceed CMB-S4 projected reach by a factor of $\mathcal{O}(2)$.\\

\item {\bf{Light Relic Particles:}} Measure the number of light particle species that were in thermal equilibrium with the known standard-model particles at any time in the early Universe, i.e.~$N_{\rm{eff}}$, with a $\sigma({N_{\rm{eff}}}) = 0.014$. This would cross the critical threshold of 0.027, which is the amount that any new particle species changes $N_{\rm{eff}}$ away from its Standard Model value of 3.04.  Such a measurement would rule out or find evidence for new light thermal particles with at least $95\%$ confidence level. This is particularly important because many dark matter models predict new light thermal particles~\cite{Baumann:2016wac}, and recent short-baseline neutrino experiments have found puzzling results possibly suggesting new neutrino species~\cite{Gariazzo2013,Gonzalez-Garcia2019}. \\

\item {\bf{Inflation:}} Probe the existence of inflationary magnetic fields in the early Universe to find evidence of inflation. The extremely tight constraint on anisotopic birefringence provided by CMB-HD will significantly tighten the constraints on the strength of scale-invariant primordial magnetic fields, $B_{\rm SI}$, that originate during inflation. CMB-HD will have the sensitivity to obtain a $1\sigma$ uncertainty of $\sigma(B_{\rm SI})=0.036~\mathrm{nG}$, which is below the $0.1\,\mathrm{nG}$ threshold that distinguishes between purely inflationary and dynamo origins of the $\mu\mathrm{G}$ level magnetic fields observed in galaxies~\cite{Mandal:2022tqu}. CMB-HD will therefore have the capability to detect inflationary PMFs with about $3\sigma$ significance or higher, or rule out a purely inflationary origin of galactic magnetic fields at over $95\%$~CL.\\

\item {\bf{Inflation:}} Measure the primordial local non-Gaussian fluctuations in the CMB, characterized by the parameter $f_\mathrm{NL}^{\rm local}$, with an uncertainty of $\sigma(f_\mathrm{NL}^{\rm local}) = 0.26$, by combining the kSZ signal from CMB-HD with an over-lapping galaxy survey such as from the Vera Rubin Observatory. Reaching a target of $\sigma(f_\mathrm{NL}^{\rm local}) < 1$ would rule out a wide class of multi-field inflation models, shedding light on how inflation happened~\cite{Alvarez:2014vva,Smith2018,Munchmeyer:2018eey,Deutsch2018,Contreras2019,Cayuso2018}. This cross-correlation could also resolve the physical nature of several statistical anomalies in the primary CMB~\cite{Cayuso2019} that may suggest new physics during inflation (see Ref.~\cite{Schwarz2015} for a review), and provide constraints on the state of the Universe before inflation~\cite{Zhang2015}.\\

\item  {\bf{Inflation:}} Remove 90\% of the CMB B-mode fluctuations from gravitational lensing leaving only 10\% remaining, i.e.~achieve $A_{\rm{lens}}=0.1$, over half the sky.  This would enable other CMB experiments with small-aperture telescopes, such as CMB-S4, or satellite missions, such as LiteBird, to achieve or improve on their target measurement of the amplitude of primordial gravitational waves, given by the parameter $r$.  In addition, CMB-HD can provide an independent constraint on primordial gravitational waves with an uncertainty of $\sigma(r)=0.005$ via polarised Sunyaev-Zel'dovich tomography. \\

\item {\bf{Dark Energy:}} Measure the dark energy equation of state with an uncertainty of $\sigma(\wde)= 0.005$ by combining galaxy cluster abundance measurements, galaxy cluster lensing measurements, and measurements of the primary CMB power spectra \citep{raghunathan21a, raghunathan21b}.  This would provide a constraint on the dark energy equation of state to sub-percent level accuracy.
\\

\item {\bf{Neutrino Mass:}} Detect the sum of neutrino masses at about $5\sigma$ significance or higher.  CMB-HD has multiple pathways to detect the sum of the neutrino masses, $\summnu$, which include combining cluster abundance and primary CMB power spectra measurements as well as combining CMB lensing and primary CMB power spectra.  The former will result in an uncertainty of $\sigma(\summnu)=13$~meV, assuming a {\it Planck}-like prior on the optical depth to reionization of $\sigma(\taure) = 0.007$ \citep{raghunathan21a, raghunathan21b}.  Given the current lower bound of $\summnu \ge 58$~meV, CMB-HD will be able to detect the sum of neutrino masses at about $5\sigma$ significance or more.
\\

\item {\bf{Other Beyond Standard Model:}} Provide a world-leading constraint on isotropic birefringence of 0.035 arcmin and on a scale-invariant birefringence power spectrum of $1.4 \times 10^{-6}$ deg$^2$ (68\% CL). High precision CMB-HD polarization data can be used to test for new physics by searching for a rotation of linear polarization as the CMB photons propagate to us from the surface of last scattering, usually referred to as cosmic birefringence. The measurement of cosmic birefringence can be used to constrain very light axion-like particles of $m_a \lesssim 10^{-28}$\,eV \cite{Harari:1992:axion,Carroll:1998,Li:2008,Pospelov:2009,Capparelli:2019:CB}, the axion string network \cite{Agrawal:2019:biref}, axion dark matter \cite{Liu:2016dcg}, general Lorentz-violating physics in the context of Standard Model extensions \cite{Leon:2017}, and primordial magnetic fields through the Faraday rotation \cite{Kosowsky:1996,Harari:1997,Kosowsky:2004:FR}. CMB-HD will improve statistical errors on the isotropic and scale-invariant anisotropic birefringence measurements by several orders of magnitude over the existing constraints \cite{P16:rot,Choi:2020ccd,Minami:2020:biref,Namikawa:2020:ACT-biref,Bianchini:2020:SPT-biref,Mandal:2022tqu}. \\

\end{enumerate}

\subsection{Astrophysics}\label{sec:astro} 
CMB-HD would also address many questions in astrophysics, such as 1.)~the evolution of gas and galaxies in the Universe.  Such a survey would also 2.)~monitor the transient sky by mapping the full observing region every few days, which opens a new window on gamma-ray bursts, novae, fast radio bursts, and variable active galactic nuclei. Moreover, CMB-HD would 3.)~provide a census of planets, dwarf planets, and asteroids in the outer Solar System, and 4.)~enable the detection of exo-Oort clouds around other solar systems, shedding light on planet formation. Finally, 5.)~CMB-HD will provide a catalog of high-redshift dusty star forming galaxies and active galactic nuclei over half the sky down to a flux limit of 0.5 mJy at 150 GHz.  The CMB-HD survey will be made publicly available, with usability and accessibility a priority. 

\section{Dark Matter and Dark Sectors}

\subsection{Gravitational Probe of Small-Scale Structure}\label{sec:grav}

Astronomical observations have provided compelling evidence for non-baryonic dark matter~\cite{Zwicky1937,Rubin1970,Ostriker1974,Fabricant1980,Bahcall1995,Clowe2006,Planck2018}. However, we have not been able to create or detect dark matter in terrestrial experiments to probe its properties directly.  If dark matter only interacts with the known standard-model Universe gravitationally, then it makes sense to explore the gravitational direction further to understand it.  An observational puzzle seems to exist regarding the distribution of matter on small-scales (scales below 10 kpc and masses below $10^9 M_{\odot}$ today); there seems to be less structure than the standard cold collisionless model of dark matter (CDM) would predict.  This may provide clues to the particle nature of dark matter, and, in fact, many well-motivated models of dark matter can explain this deviation from CDM~\cite{Colin2000,Bode2001,Boehm:2001hm,Boehm:2004th,Viel2005,Turner1983,Press1990,Sin1994,Hu2000,Goodman2000,Peebles_2000,Amendola2006,Schive2014,Marsh2016,Carlson1992,Spergel2000,Vogelsberger2012,Dvorkin:2013cea,Fry2015,Elbert2015,Kaplinghat2016,Kamada2017,Gluscevic2018,Boddy2018a,Li2018,Boddy2018b,Tulin2018,Khoury2016,Vogelsberger2016,Cyr-Racine:2013ab,Cyr-Racine:2015ihg,Schewtschenko:2014fca,Krall:2017xcw,Xu:2018efh,2019ApJ...878L..32N,2021PhRvD.104j3521N,2021ApJ...907L..46M}.

However, measurements of the small-scale matter distribution do not conclusively indicate a deviation from the CDM prediction.  This is because such measurements often infer the matter distribution through baryonic tracers~\cite{Koposov2015,Drlica-Wagner2015,Menci2017,Mesinger2005,deSouza2013,Moore1999,Johnston2016,Carlberg2009,Erkal2015,Bovy2014,Cen1994,Hernquist1996,Croft1999,Hui1999,Viel2013,Baur2016,Irsic2017}, and such tracers may not reliably map the dark matter~\cite{Sawala2016,Oman2015,Hui2017}. Gravitational lensing offers a powerful way to map the dark matter directly. While strong gravitational lensing is a promising method to find low-mass dark-matter halos~\cite{Dalal2002,Koopmans_2005,Vegetti:2009aa,Vegetti_2010_1,Vegetti_2010_2,Vegetti2012,Hezaveh:2012ai,Vegetti2014,Hezaveh2016a,Ritondale:2018cvp}, it does face the challenge of separating the complex and often unknown structure of the background source from the sought-after substructure signal; using strong lensing to measure the matter power spectrum, faces a similar challenge~\cite{Hezaveh2016b,Daylan:2017kfh,Bayer:2018vhy,DiazRivero2018,Chatterjee2018,Cyr-Racine:2018htu,Brennan2018,Rivero:2018bcd}.  A method that can evade this challenge is to measure the small-scale matter power spectrum from weak gravitational lensing using the CMB as a backlight~\cite{Nguyen2019}.  The CMB serves as a perfect backlight because i)~it has a known redshift, ii)~is behind every object, and iii)~is a known pure gradient on the small-scales of interest. Thus this method can provide a powerful complementary probe of dark-matter physics~\cite{Nguyen2019}. 

\begin{figure}[t]
\centering
\includegraphics[width=0.47\textwidth]{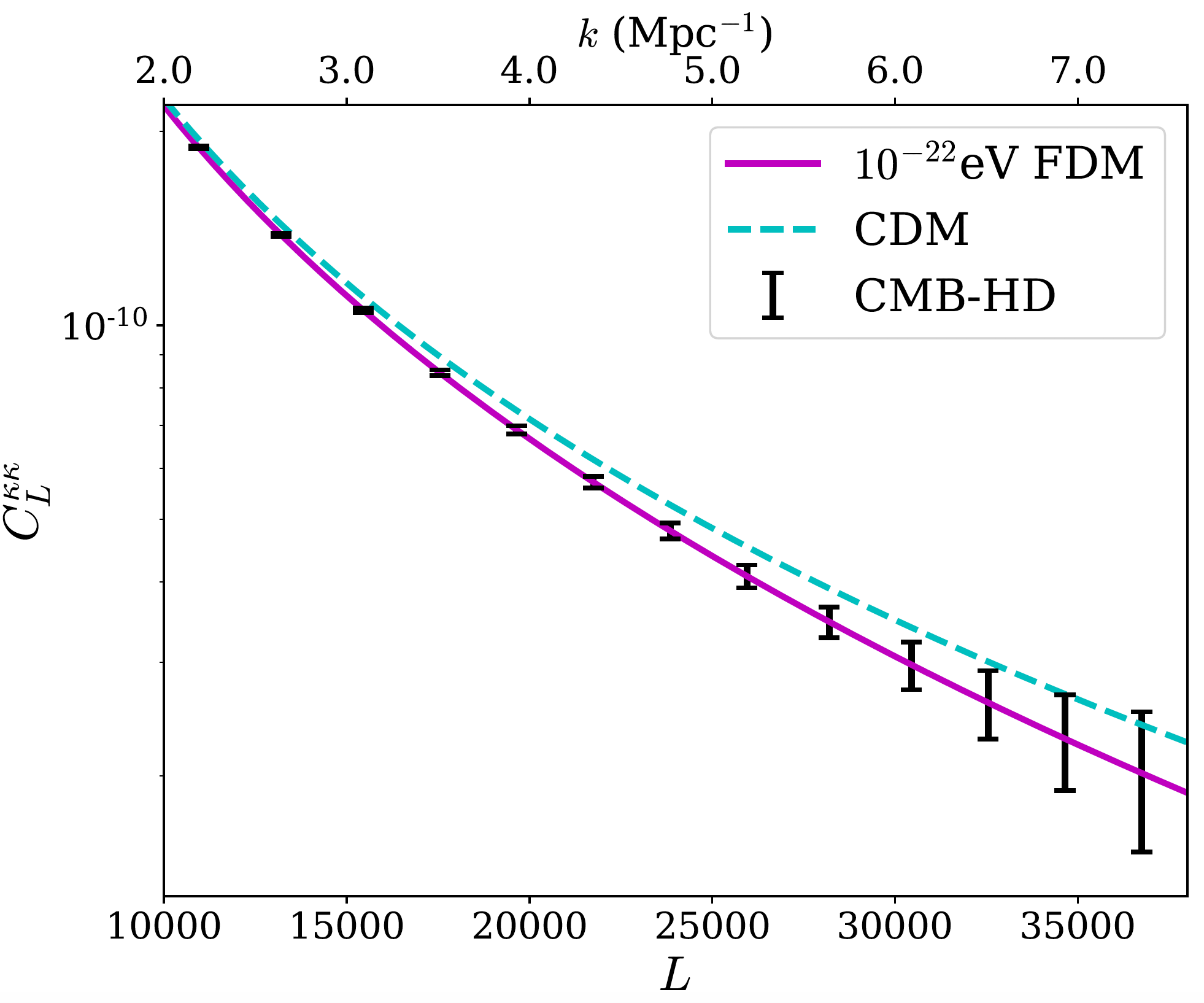}\hfill\includegraphics[width=0.5\textwidth,height=6.1cm]{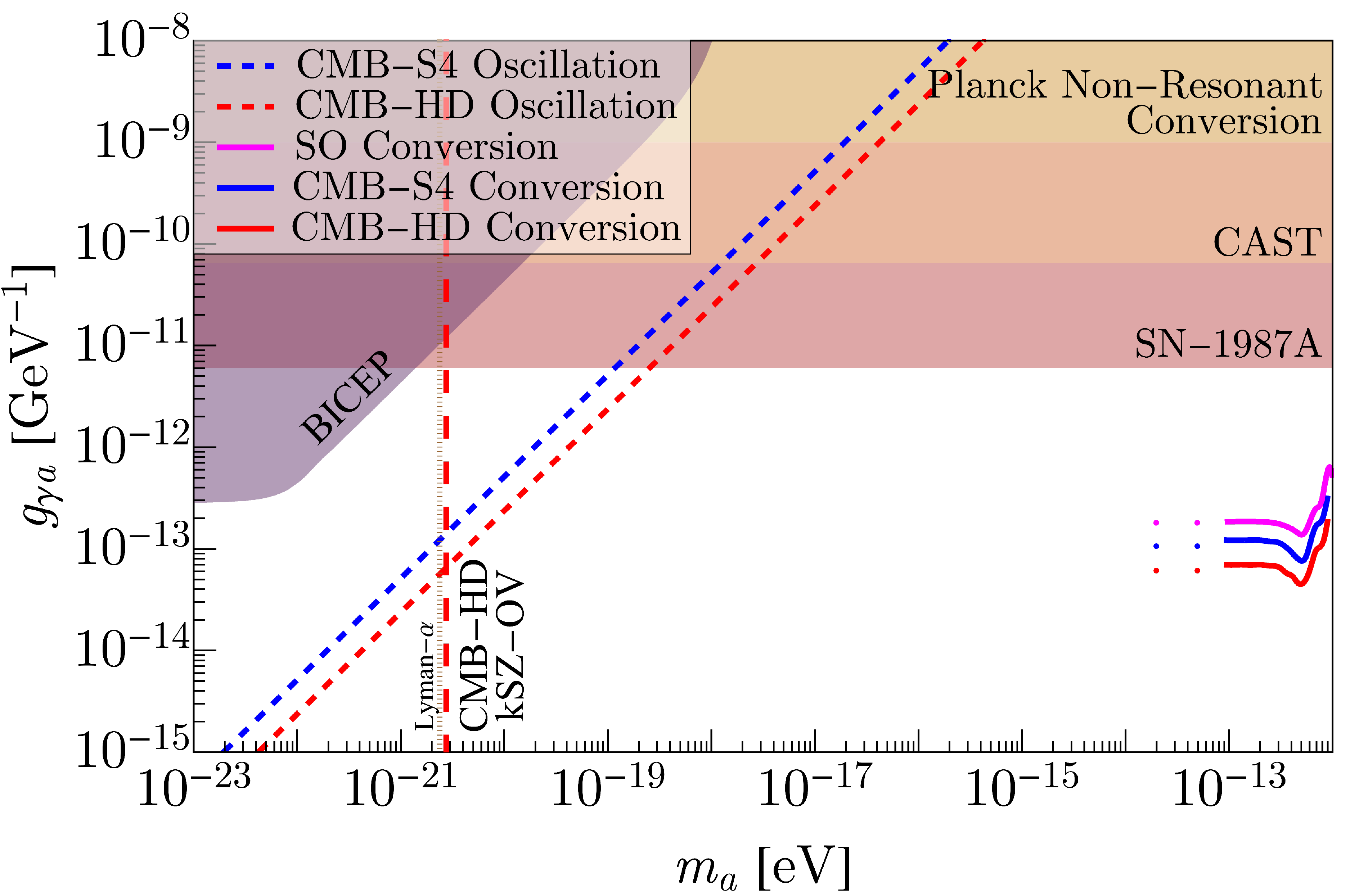}
\caption{{\it{Left:}} {\bf{Dark Matter:}} CMB-HD would generate a high-resolution map out to $k\sim10~h$Mpc$^{-1}$ of the projected dark matter distribution over half the sky via gravitational lensing. Shown is the CMB lensing power spectrum for an $m \sim 10^{-22}$ eV FDM model and a CDM model.  Error bars correspond to CMB-HD expectations of 0.5 $\mu$K-arcmin noise in temperature and 15 arcsecond resolution over $50\%$ of the sky. {\it{Figure credit: \foreignlanguage{vietnamese}{Hồ Nam Nguyễn}}.} {\it{Right:}} {\bf{Axion-like Particles:}} Forecasted constraints on axion-like particles from the resonant conversion of CMB photons into axions-like particles in the magnetic fields of galaxy clusters, from oscillation of the local CMB polarization angle, and from the kinetic SZ Ostriker-Vishniac (kSZ-OV) sensitivity to the mass of a very light ALP (using the methods of~\cite{Farren:2021jcd}), comparable to constraints from Lyman-$\alpha$ forest (e.g.~\cite{Rogers:2020ltq}).  The BICEP curve is obtained from Eq.~(15) of~\cite{BICEPKeck:2021sbt}, assuming $\kappa\rho_0=0.3\,\mathrm{GeV\,cm^{-3}}$. {\it{Figure credit: Suvodip Mukherjee, Michael Fedderke, Sayan Mandal, Gerrit Farren, and Daniel Grin.}}}
\label{fig:DMandAxions}  
\end{figure}

We show in the left panel of Figure~\ref{fig:DMandAxions} how one can distinguish between CDM and a model of dark matter that suppresses structure on small scales, by measuring the small-scale gravitational lensing of the CMB~\cite{Nguyen2019}. 
Including foregrounds, one can potentially distinguish between CDM and a 1 keV warm dark matter model or a $10^{-22}$~eV fuzzy dark matter (FDM) model at the $5\sigma$ level~\cite{Han:2021vtm}. Such observations could also probe scattering interactions between dark matter and a new relativistic component \cite{Krall:2017xcw} or the scattering rate of self-interacting dark matter \cite{Li:2018zdm}.
Newer lensing reconstruction techniques can potentially improve these constraints further~\cite{Horowitz2019,Hadzhiyska2019}. Extragalactic foregrounds are the main source of systematic effect in this measurement, and paths to mitigate this are discussed in~\cite{Han:2021vtm} and in Section~\ref{sec:fg}.  Baryonic processes can also move around the dark matter and change the shape of the small-scale matter power spectrum; however, they likely affect the shape of the spectrum in a way that differs from alternate dark matter models~\cite{Nguyen2019,vanDaalen2011,Brooks2013,Brooks2014,Natarajan2014,Schneider2018}.  Comparing the matter power spectrum of multiple hydrodynamic simulations, each with different baryonic prescriptions, suggests there may be a finite set of ways that baryons can change the shape of the spectrum, characterized by just a few free parameters~\cite{Schneider2018}. Given that, one can use the shape of the power spectrum to distinguish between dark matter models and baryonic effects.  In any case,this measurement would be a clean measurement of the matter power spectrum on these scales, free of baryonic tracers. It would greatly limit the allowed models of dark matter and baryonic physics, shedding light on dark-matter properties as well as galaxy evolution. \\

\subsection{Resonant Photon-ALPs Conversion}
\label{sec:axion-like1}

The existence of axion or axion-like particles (ALPs) in nature is a prediction from theoretical physics \cite{PhysRevLett.38.1440, PhysRevLett.40.223, PhysRevLett.40.279, Svrcek:2006yi, Arvanitaki:2009fg} and its discovery is going to be a paradigm shift in the framework of theoretical physics. Several particle physics experiments  such as CERN ALP Solar Telescope (CAST) \cite{2015PhPro..61..153R}, LPS-II \cite{Bastidon:2015efa}, MADMAX \cite{Majorovits:2016yvk}, ADMX \cite{2010PhRvL.104d1301A}, CASPER \cite{Budker:2013hfa} are looking for the signatures of ALPs over a wide range of masses. Along with the particle physics experiments, cosmological probes such as CMB anisotropies and large-scale structure are exploring the gravitational effect of ALPs on the matter density, and this is a potential probe to discover ALPs if they are the dark matter, as discussed in Section~\ref{sec:grav}~\cite{Hu:2000ke, Hlozek:2014lca,Nguyen2019}. Other possibilities to probe ALPs via cosmological observables, in some cases even if ALPs are not the dark matter, are through their coupling with photons.

One possibility to study ALPs is to use their coupling with photons in the presence of an external magnetic field \cite{PhysRevLett.51.1415, PhysRevD.37.1237}. The coupling between ALPs and photons denoted by $g_{\gamma\gamma a}$ (we will use the notation $g_{\gamma a}$ throughout) leads to oscillations between photons and ALPs and vice versa in the presence of an external magnetic field. This effect is one of the cleanest windows to detect ALPs irrespective of whether it is dark matter or not. The signatures of this non-gravitational interaction of ALPs with photons distorts the energy spectrum of photons and can be detected robustly from a source of radiation with a well-known energy spectrum. The radiation field of the CMB provides us with an excellent source that can be used to detect the distortions due to ALPs \cite{Mukherjee:2018oeb, Mukherjee:2018zzg, Mukherjee:2019dsu}. The  ALP distortion ($\alpha$-distortion) gets imprinted on the CMB while it is passing through the external magnetic field of the intergalactic medium (IGM) \cite{Mukherjee:2018oeb}, inter-cluster medium (ICM) \cite{Mukherjee:2019dsu}, voids \cite{Mukherjee:2018oeb} and Milky Way (MW) \cite{Mukherjee:2018oeb}. The conversion from photons to ALPs can be classified into two types, namely the resonant conversion and the non-resonant conversion. 

The resonant conversion of CMB photons into ALPs takes place when the effective photon mass in the plasma equals the mass of the ALP. The polarization state of the CMB photon that is parallel to the external magnetic field gets converted into ALPs depending upon the strength of the magnetic field. As a result, it leads to a polarized spectral distortion in the CMB blackbody with a unique spectral shape. Also due to inhomogeneities in the magnetic field in astrophysical systems, the observed polarized distortion varies spatially, which leads to a unique spatial structure that is different from any other known spectral distortions and foreground contamination. Though the resonant conversion of CMB photons can take place in different kinds of astrophysical systems, it can be best measured in the Milky Way \cite{Mukherjee:2018oeb} and galaxy clusters \cite{Mukherjee:2019dsu}. More details regarding the resonant photon-ALPs spectral distortion in Milky Way and galaxy clusters can be found in these works \cite{ Mukherjee:2018oeb, Mukherjee:2018zzg, Mukherjee:2019dsu}. 

In the right panel of Figure~\ref{fig:DMandAxions}, we show the expected constraints on the photon-ALP coupling strength as a function of ALP mass for CMB-HD (red solid line). For comparison, we have plotted constraints from upcoming CMB missions such as the Simons Observatory \cite{Ade:2018sbj} and CMB-S4 \cite{Abazajian:2016yjj}. For masses smaller than $\sim 10^{-13}$ eV, the constraints are shown as dotted lines since they depend on the availability of magnetic field observations in the outskirts of galaxy clusters; it is possible to achieve this by combining with magnetic field measurements using radio observations. We also show the current bounds on photon-ALP coupling from the Planck temperature map \cite{Mukherjee:2018zzg},  CAST \cite{Anastassopoulos:2017ftl}, and SN1987A \cite{2015JCAP...02..006P}.

\subsection{CMB Polarization Oscillation}
\label{sec:axion-like2}

Another possibility to study ALPs via the ALP-photon coupling $\mathcal{L} \supset - \frac{1}{4} g_{a\gamma}aF_{\mu\nu}\tilde{F}^{\mu\nu}$ arises when linearly polarized light traverses a region in which there is a varying ALP background field; in this case, this coupling generates a birefringence effect for photons that manifests as a rotation of the linear polarization of the light~\cite{Carroll:1989vb,Harari:1992:axion,Carroll:1998}. Since the ALP-photon coupling is a derivative interaction, the integrated angular amplitude of this rotation effect $\Delta \Psi$ depends only on the values of the ALP field at the end-points of the light trajectory~\cite{Harari:1992:axion,Ivanov:2018byi,Fedderke:2019ajk}: $\Delta \Psi = \frac{1}{2} g_{a\gamma} \left( a_{\mathrm{det}} - a_{\mathrm{emit}} \right)$, where $a_{\mathrm{emit, det}}$ are, respectively, the ALP field values at the spacetime points where the light is emitted and detected. In the case where the value of the ALP field changes slowly on cosmologically slow timescales, this is the origin of the well-known ALP-induced cosmic birefringence effect~\cite{Carroll:1989vb,Harari:1992:axion,Carroll:1998}. 

However, assuming the ALP is all of the dark matter, and assuming it is constrained by small-scale structure observables~\cite{Irsic:2017yje,DES:2020fxi,2019ApJ...878L..32N} to have a mass $m_a \gtrsim 2\times 10^{-21}\,\textrm{eV}$ (somewhat stronger bounds have also appeared~\cite{Rogers:2020ltq}), then since the non-relativistic ALP dark matter field oscillation period is uniquely fixed by its mass to be $T_{a} = 2\pi\hbar / m_a$, it oscillates on timescales no longer than $\sim$ a month. This is extremely rapid by cosmological standards and gives rise to two novel effects on the CMB polarization pattern which go beyond the usual cosmic birefringence effect~\cite{Fedderke:2019ajk}: a partial depolarization of the CMB driven by ALP dark matter field oscillations around the time of CMB decoupling, and a time-dependent oscillation of the local CMB polarization angle as measured on the sky today driven by the oscillation of the local ALP dark matter field at the location of the detector. The partial depolarization (or `washout') effect has been examined in~\cite{Fedderke:2019ajk} using Planck power spectrum data and shown to constrain $g_{a\gamma} \lesssim 10^{-12}\,\textrm{GeV}^{-1} \times (m_a / 10^{-21}\,\textrm{eV})$; this effect is however limited by cosmic variance and can be improved beyond these existing limits by at most a factor of $\mathcal{O}(3)$.

On the other hand, the oscillation effect (`AC oscillation') holds much more promise~\cite{Fedderke:2019ajk}. The effect can be phrased as a local-on-the-sky oscillation of the \emph{measured} $Q,U$ patterns~\cite{Fedderke:2019ajk,BICEPKeck:2020hhe,BICEPKeck:2021sbt} or, equivalently, an oscillation of the \emph{measured} CMB polarization angle. Because the rotation of the polarization angle is caused directly by the oscillation of the local ALP dark matter field value, its frequency is fixed uniquely by the ALP mass (a fundamental physics parameter), and it has a very narrow Fourier space linewidth: $\Delta f / m_a \sim v_{\textsc{dm}}^2 \sim 10^{-6}$. This narrowband property guarantees that, for ALPs with $m_a \lesssim 10^{-18}\,\textrm{eV}$ (i.e., periods $\gtrsim$ hours), the rotation is phase-coherent on timescales that exceed 100\,yrs, which also vastly exceeds observational timescales; the signal sensitivity thus builds coherently with integration time, and can be compared phase-coherently between detectors operating at different epochs. The coherence length of the ALP dark matter field is also much larger than the Solar System at these ALP masses, so the rotation can be also phase-coherently compared between different detectors operating at different locations on Earth or in orbit. Moreover, the phase of the rotation of the CMB polarization angle does not depend on the direction of arrival of the light at the detector: the signal is in-phase across the entire sky. Finally, because the signal is a temporal variation of the \emph{measured} $Q,U$ patterns, it is not limited by cosmic variance. CMB-HD thus holds enormous potential for extending the reach of this possible ALP dark matter discovery channel. 

Assuming that no ALP dark matter detection occurs, the sensitivity of CMB-HD to exclude ALP dark matter parameter space via a search for narrowband temporal fluctuation of the CMB polarization pattern with periods of $\sim$ hours to $\sim$ months can be roughly estimated from the statistical power of the polarization maps as follows.
The amplitude of the $Q,U$ oscillation due to the rotation effect is of order~\cite{Fedderke:2019ajk,BICEPKeck:2020hhe,BICEPKeck:2021sbt} $\Delta \{Q,U\} \sim g_{a\gamma}a_0 \{U,Q\}$, where $a_0 \sim \sqrt{2\rho_{\textsc{dm}}}/m_a$ is the ALP field amplitude assuming that the ALP constitutes all of the local dark matter, $\rho_{\textsc{dm}}\sim 0.3\,\textrm{GeV/cm}^3$. The size of this variation should be compared to the statistical error on the full-map measurement uncertainty of $Q,U$, which is $\delta_{Q,U} \sim \sigma_{\textrm{pol}}/\sqrt{\Omega_{\textrm{map}}}$ with $\sigma_{\textrm{pol}}$ the polarization map depth in $\mu$K-arcmin and $\Omega_{\textrm{map}}$ the sky area mapped in (arcmin)${}^2$; we assume that the noise is white. 
Using that the rms values of the $Q,U$ fields over the anticipated CMB-HD map area are similar ($Q_{\textrm{rms}}\sim U_{\textrm{rms}}$)~\cite{Sehgal:2020abd}, and accounting for various $\mathcal{O}(1)$ factors, this comparison leads to a projection for the statistical sensitivity to exclude ALP dark matter parameter space at 95\%-confidence (ignoring systematics): $g_{a \gamma}\lesssim g^{95}_{a \gamma} \sim m_a \times \big(\sqrt{2}\sigma_{\textrm{pol}}\big)/\big(Q_{\textrm{rms}} \sqrt{\Omega_{\textrm{map}} \cdot \rho_{\textsc{dm}}} \big)$.
This rough estimate is a factor of $\mathcal{O}(3)$ stronger than existing detailed BICEP/Keck analyses when the comparison is made using appropriate parameters~\cite{BICEPKeck:2020hhe,BICEPKeck:2021sbt}, and should be considered to be accurate at that level.
Numerically, the search sensitivity assuming fiducial CMB-HD parameters is
\begin{equation}
    g^{95}_{a \gamma} \sim  
    2.1\times 10^{-14}\,\textrm{GeV}^{-1} \times \left( \frac{m_{a}}{10^{-21}\,\textrm{eV}} \right) \times \left( \frac{ \sigma_{\textrm{pol}}}{ 0.76\,\mu\textrm{K-arcmin}} \right) \times \left( \frac{ f_{\textrm{sky}}}{0.5} \right)^{-1/2}\times \left( \frac{Q_{\textrm{rms}}}{ 4\,\mu\textrm{K} } \right)^{-1},
\end{equation}
where $f_{\textrm{sky}}$ is the fractional sky coverage.
The fiducial value for the polarization map-depth shown above is the combined 90/150~GHz sensitivity for CMB-HD, and we have assumed the appropriate half-sky-averaged rms value for the $Q,U$ maps~\cite{Sehgal:2020abd}. This sensitivity projection is shown in the right panel of Figure \ref{fig:DMandAxions}, and is a couple of orders of magnitude greater than that of existing BICEP/Keck analyses searching for this effect~\cite{BICEPKeck:2020hhe,BICEPKeck:2021sbt}.
The cognate estimate for CMB-S4 combining both narrow- and wide-field components~\cite{Abazajian:2019eic} yields a search sensitivity reduced from that of CMB-HD by a factor of $\mathcal{O}(2)$.

\section{Light Relic Particles}
\subsection{Free-streaming Relativistic Species}

The CMB can tell us about the inventory of particles that existed in the Universe right back to the Big Bang.  This is because any light particles (with masses less than 0.1 eV) that were in thermal equilibrium with the known standard-model particles, at any time in the early Universe, would have left an imprint on the CMB. Even one new species of particle as described above, would change the effective number of light particles, $N_{\rm{eff}}$, by as much as $\sigma({N_{\rm{eff}}}) = 0.027$ away from the standard model value (assuming no significant dilution by new states beyond the standard model particle content)~\cite{CMBS4SB}. 
If these light particles interact only very weakly, we may never see them in laboratory experiments, and astronomical measurements may provide the only avenue for their detection.
This is particularly important because many dark matter models predict new light thermal particles~\cite{Baumann:2016wac,Green:2017ybv}, and some short-baseline neutrino experiments have found puzzling results possibly suggesting new neutrino species~\cite{Gariazzo2013,Gonzalez-Garcia2019}.

CMB-HD can achieve $\sigma({N_{\rm{eff}}}) = 0.014$, which would cross the critical threshold of 0.027.  This would potentially rule out or find evidence for new light thermal particles with $95\%$ confidence level. By combining the CMB-HD primordial CMB measurements with the CMB-HD measurement of the small-scale matter power spectrum discussed above, we can also potentially gain a factor of two improvement on $\sigma({N_{\rm{eff}}})$, subject to improvements in modelling the nonlinear matter power up to $k\sim0.4~h$Mpc$^{-1}$ \cite{Baumann2018}. Removal of foregrounds is important for achieving these constraints, and we discuss paths to do so in Section~\ref{sec:fg}; we note here that the high-angular resolution and multiple frequency channels of CMB-HD, spanning 30 to 350 GHz, help make the necessary foreground cleaning possible. \\

\subsection{Fluid-like and Self-interacting Relativistic Species}

We might relax the assumption that light relics only interact gravitationally, i.e.~that they are no longer free-streaming. While free-streaming light relics induce a phase shift of the acoustic peaks in primary CMB observations~\citep{Bashinsky:2003tk,Baumann:2015rya}, sufficiently strong interactions can counteract this effect, and additionally cause an increase in the amplitude of the temperature and polarization spectra, and a scale-dependent modulation of the CMB lensing amplitude~\citep{Kreisch:2019yzn,Brinckmann:2020bcn}. The latter effect is a particularly promising target for CMB-HD, as the signal increases towards smaller scales, which are accessible for CMB-HD. As such, it is expected that CMB-HD can place more stringent constraints on the effective number of relativistic species when allowing for interactions, e.g.~in the form of fluid-like, decoupling, or recoupling phenomenologies from neutrino self-interactions, neutrino-dark sector interactions, or dark radiation self-interactions~\citep{Bell:2005dr,Friedland:2007vv,Cyr-Racine:2013jua,Oldengott:2014qra,Wilkinson:2014ksa,Escudero:2015yka,Buen-Abad:2015ova,Chacko:2016kgg,Buen-Abad:2017gxg,Brust:2017nmv,Choi:2018gho,Escudero:2019gvw,Ghosh:2019tab,Blinov:2020hmc,Das:2020xke,Escudero:2021rfi,Aloni:2021eaq}. Some of these models may have the potential to alleviate the Hubble tension~\citep{Schoneberg:2021qvd}.

\begin{figure}[t]
\centering
\includegraphics[width=0.5\textwidth,height=6.1cm]{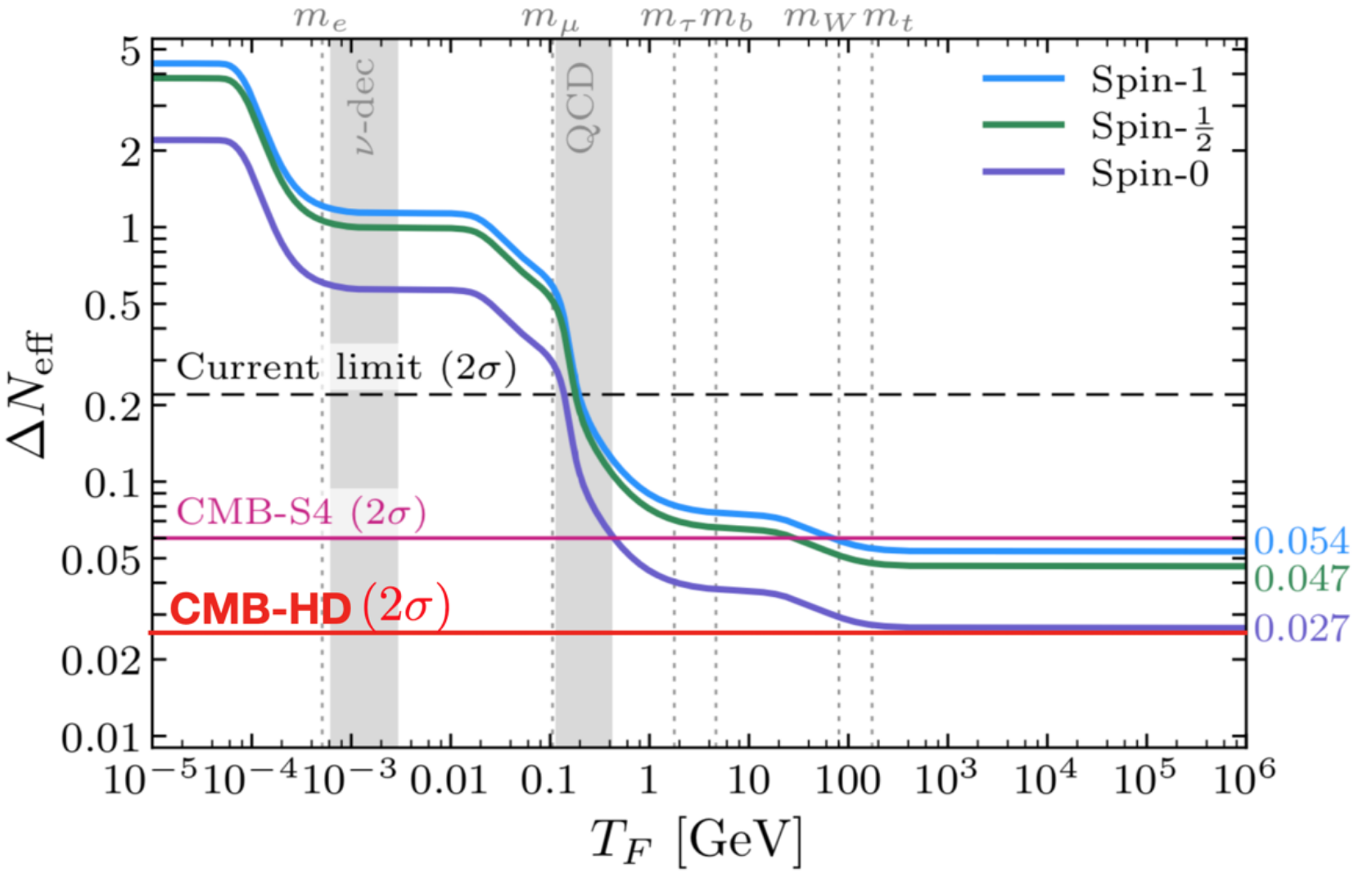}\hfill\includegraphics[width=0.47\textwidth]{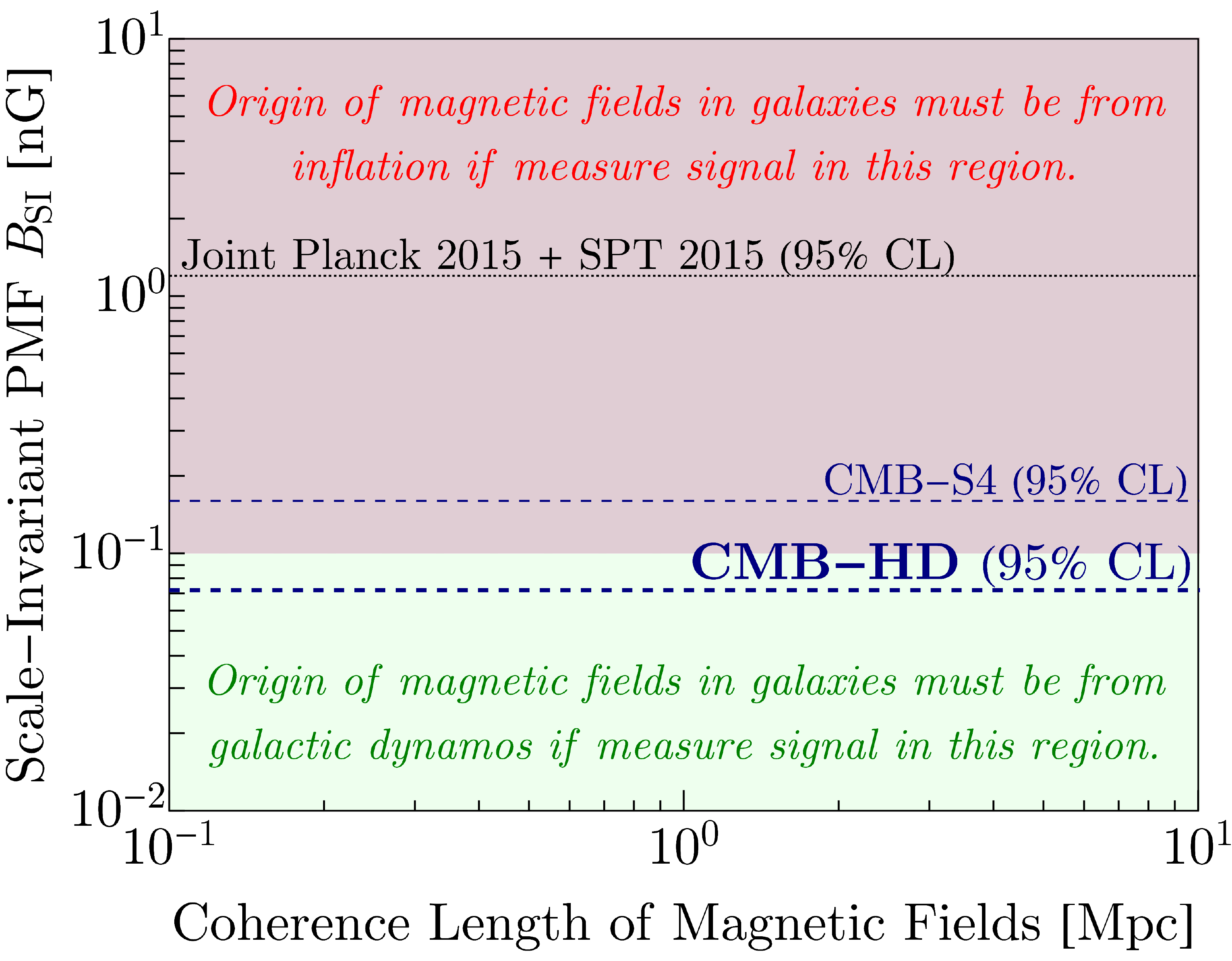}
\caption{{\it{Left:}} {\bf{Light Relic Particles:}} CMB-HD can achieve $\sigma({N_{\rm{eff}}}) = 0.014$, which would cross the critical threshold of 0.027, ruling out or finding evidence for new light thermal particles, at any time in the early Universe, with at least $95\%$ confidence level.  {\it{Original figure by Benjamin  Wallisch from~\cite{Green2019,Wallisch2018}; modified with addition of CMB-HD forecast.}} {\it{Right:}} {\bf{Inflation:}} We show the forecasted $95\%$ CL upper bounds on $B_{\rm SI}$ from anisotropic birefringence measurements for CMB-HD (thick-dashed lines) and also CMB-S4~\cite{Mandal:2022tqu}. We compare it with the current $95\%$ CL upper bound on $B_{\rm SI}$ (dotted black line) obtained from CMB temperature and polarization spectra by a joint Planck and SPT analysis~\cite{Pogosian:2019:axion}. {\it{Figure credit: Sayan Mandal.}}}
\label{fig:NeffandPMF}  
\end{figure}

\section{Inflation}

\subsection{Inflationary Magnetic Fields}
\label{sec:PMFs}

The origin of the $\mathcal{O}(\mu\mathrm{G})$ magnetic fields observed in galaxies today is among the biggest unsolved problems in astrophysics~\cite{Widrow:2002ud, Durrer:2013pga}.
These magnetic fields could have originated in either of the following processes:

\begin{enumerate}
    \item during \textbf{inflation} or related processes such as reheating or preheating from the amplification of quantum vacuum fluctuations~\cite{Widrow:2002ud, Ratra:1991bn}, leading to scale-invariant magnetic fields,
    \item during \textbf{phase transitions} in the early Universe, like the electroweak (EW) or quantum chromodynamics (QCD) phase transitions, where a strong first-order phase transition is required to generate magnetic fields~\cite{Hogan:1983zz, Witten:1984rs}, or
    \item from weak seed fields or local plasma mechanisms in galaxies which are amplified by \textbf{galactic dynamo processes}~\cite{Kulsrud:2007an}.
\end{enumerate}

Magnetic fields generated by the first two processes are called \textit{primordial magnetic fields} (PMFs)~\cite{Kandus:2010nw}; in this scenario, the $\mathcal{O}(\mu\mathrm{G})$ magnetic fields we observe in galaxies today are a result of adiabatic compression during structure formation of $\mathcal{O}(\mathrm{nG})$ PMFs present on Mpc scales after recombination.
PMFs are also an attractive way to explain the presence of weak magnetic fields, of at least $10^{-6}\,\mathrm{nG}$ on Mpc scales, in the empty voids between galaxy clusters~\cite{Neronov:2010gir, Fermi-LAT:2018jdy, VERITAS:2017gkr}, since these fields are observed to be relatively uniformly spread within the voids.

PMFs before structure formation are sustained in the ionized plasma, which later adiabatically collapses to form galaxies.
PMFs of strength above $0.1\,\mathrm{nG}$ would lead to $\mu\mathrm{G}$ magnetic fields in galaxies due to flux conservation~\cite{Grasso:2000wj}, i.e.~$~0.1\,\mathrm{nG} =  1\,\mu\mathrm{G} (10~{\rm kpc} / 1~{\rm Mpc})^2$, assuming a conservative characteristic radius of baryonic matter in galaxies of 10 kpc.
Since phase transitions do not generate scale-invariant PMFs, detecting scale-invariant PMFs ($B_{\rm SI}$) above $\sim 0.1\,\mathrm{nG}$ on $\mathrm{Mpc}$ scales post-recombination would imply that the galactic magnetic fields originate during inflation~\cite{Grasso:2000wj, Widrow:2011hs, Ryu:2011hu}. Moreover, detecting such inflationary PMFs would be a smoking-gun signature of inflation itself since such a strong scale-invariant PMF on Mpc scales can only be generated during inflation~\cite{Durrer:2013pga, Subramanian:2015lua}.
On the other hand, if we constrain $B_{\rm SI}<0.1\,\mathrm{nG}$, the galactic fields could not have originated purely from inflation, and would require further amplification from galactic dynamo processes.

The strength of scale-invariant PMFs on Mpc scales can be constrained through their imprint on the CMB~\cite{Pogosian:2018vfr}.
PMFs induce temperature and polarization anisotropies in the CMB~\cite{Durrer:2006pc, Subramanian:2006xs, Subramanian:2015lua}, affecting their respective power spectra.
PMFs present after recombination also cause anistropic birefringence (or Faraday Rotation) of the CMB, i.e., a rotation of the plane of polarization of CMB photons~\cite{Kosowsky:1996, Kosowsky:2004:FR, Campanelli:2004pm}.
Currently, scale-invariant PMFs are constrained to be below $1.2\,\mathrm{nG}$ at $95\%$ CL from a combination of the Planck TT, EE, and TE spectra and the BB spectrum from SPT~\cite{Pogosian:2019:axion}.
Since anisotropic birefringence scales as the square of the magnetic field strength~\cite{Durrer:2013pga, Subramanian:2015lua}, as opposed to the fourth power for the CMB spectra, the vastly improved anisotropic birefringence measurement from CMB-HD (see Sec.~\ref{sec:Biref}) will lead to the tightest bound on $B_{\rm SI}$.

Using a realistic power spectrum of PMFs that accounts for MHD turbulence prior to recombination, one can calculate the expected anisotropic birefringence signal from scale-invariant PMFs in order to forecast constraints on $B_{\rm SI}$ from birefringence measurements~\cite{Mandal:2022tqu}.
The forecasted $95\%$ CL upper limit on $B_{\rm SI}$ from anisotropic birefringence for CMB-HD is shown in the right panel of Fig.~\ref{fig:NeffandPMF}, and compared to the forecast for CMB-S4 and the current $95\%$ CL upper bound of $1.2\,\mathrm{nG}$~\cite{Pogosian:2019:axion}.
Anisotropic birefringence measurements from CMB-HD will achieve a $1\sigma$ uncertainty of $\sigma(B_{\rm SI})=0.036,\mathrm{nG}$, which is well below the $0.1\,\mathrm{nG}$ threshold distinguishing between a purely inflationary origin of galactic magnetic fields and a scenario requiring amplification by galactic dynamos.
Ruling out Mpc scale PMFs above $0.1\,\mathrm{nG}$ will rule out inflation as the sole origin of the galactic magnetic fields observed today, and a \textit{detection} of PMFs below $0.1\,\mathrm{nG}$ would imply an origin through galactic dynamo processes.
Conversely, \textit{detecting} $B_{\rm SI}$ above $0.1\,\mathrm{nG}$ will imply inflationary PMFs are responsible for galactic magnetic fields, and will provide compelling evidence for inflation itself.
CMB-HD will have the capability to detect inflationary PMFs at about the $3\sigma$ level or higher, or rule out a purely inflationary origin of galactic magnetic fields at over $95\%$~CL.

\subsection{Primordial Non-Gaussianity}
\label{sec:fNL}

Primordial non-Gaussianity, the departure of the primordial curvature perturbations from Gaussianity, is a key observational target to discriminate different models of inflation (and its alternatives). Higher-order correlations of the primordial fluctuations are directly sensitive to the dynamics and field content of the primordial universe. While primordial non-Gaussianity is very small for conventional single-field slow-roll inflation models, other theoretically attractive models predict signals that could be observable with CMB-HD. In particular, so called local non-Gaussianity, quantified by the parameter $f_\mathrm{NL}^{\rm local}$, can detect whether light degrees of freedom other than the inflaton contribute to the observed scalar fluctuations. A theoretically well motivated observational target is to constrain $f_\mathrm{NL}^{\rm local} < 1$ \cite{Alvarez:2014vva}. A detection of $f_\mathrm{NL}^{\rm local} \simeq 1$ or larger requires multifield dynamics, while $f_\mathrm{NL}^{\rm local}  \ll 1$ favours single-field inflation. The current leading bound, from Planck, is $f_\mathrm{NL}^{\rm local} = -0.9 \pm 5.1 $ \cite{Akrami:2019izv}. 

CMB-HD will contribute to measuring primordial non-Gaussianity not by measuring primary CMB perturbations, which will have been measured to their cosmic variance limit in both $T$ and $E$ below $\ell = 2000$ by other experiments, but rather indirectly through secondary anisotropies. CMB-HD will improve constraints on $f_\mathrm{NL}^{\rm local}$ by exploiting CMB anisotropies from lensing and from the kinetic Sunyaev-Zeldovich (kSZ) effect. These secondary anisotropies probe non-Gaussianity in cross-correlation with a galaxy survey through scale-dependent bias \cite{Dalal:2007cu} of the galaxy field and sample variance cancellation \cite{Seljak:2008xr} with the CMB anisotropies. This technique has been developed for CMB lensing in \cite{Schmittfull:2017ffw} and for kSZ anisotropies in \cite{Munchmeyer:2018eey}, with the kSZ providing somewhat stronger constraints. Because of its unprecedented mapping of secondary CMB anisotropies, CMB-HD will be the ultimate CMB experiment for this analysis. Using the kSZ forcasting pipeline developed in \cite{Munchmeyer:2018eey}, we find that CMB-HD together with the LSST-Y10 gold sample \cite{2009arXiv0912.0201L} from the Rubin Observatory can constrain $f_\mathrm{NL}^{\rm local}$ to a precision of $\sigma({f_\mathrm{NL}^{\rm local}})=0.26$, excluding $f_\mathrm{NL}^{\rm local}=1$ at $4\sigma$ significance.  In contrast, CMB-S4 plus LSST can achieve $\sigma({f_\mathrm{NL}^{\rm local}})=0.7$~\cite{Munchmeyer:2018eey}. This constraint is limited by the galaxy sample from the Rubin Observatory, rather than by CMB-HD, and a combination with future more high-resolution galaxy surveys would lead to even better constraints.

\subsection{Primordial Gravitational Waves}
\label{sec:PGW}

\begin{figure}[t!]
\centering
\includegraphics[width=0.6\textwidth]{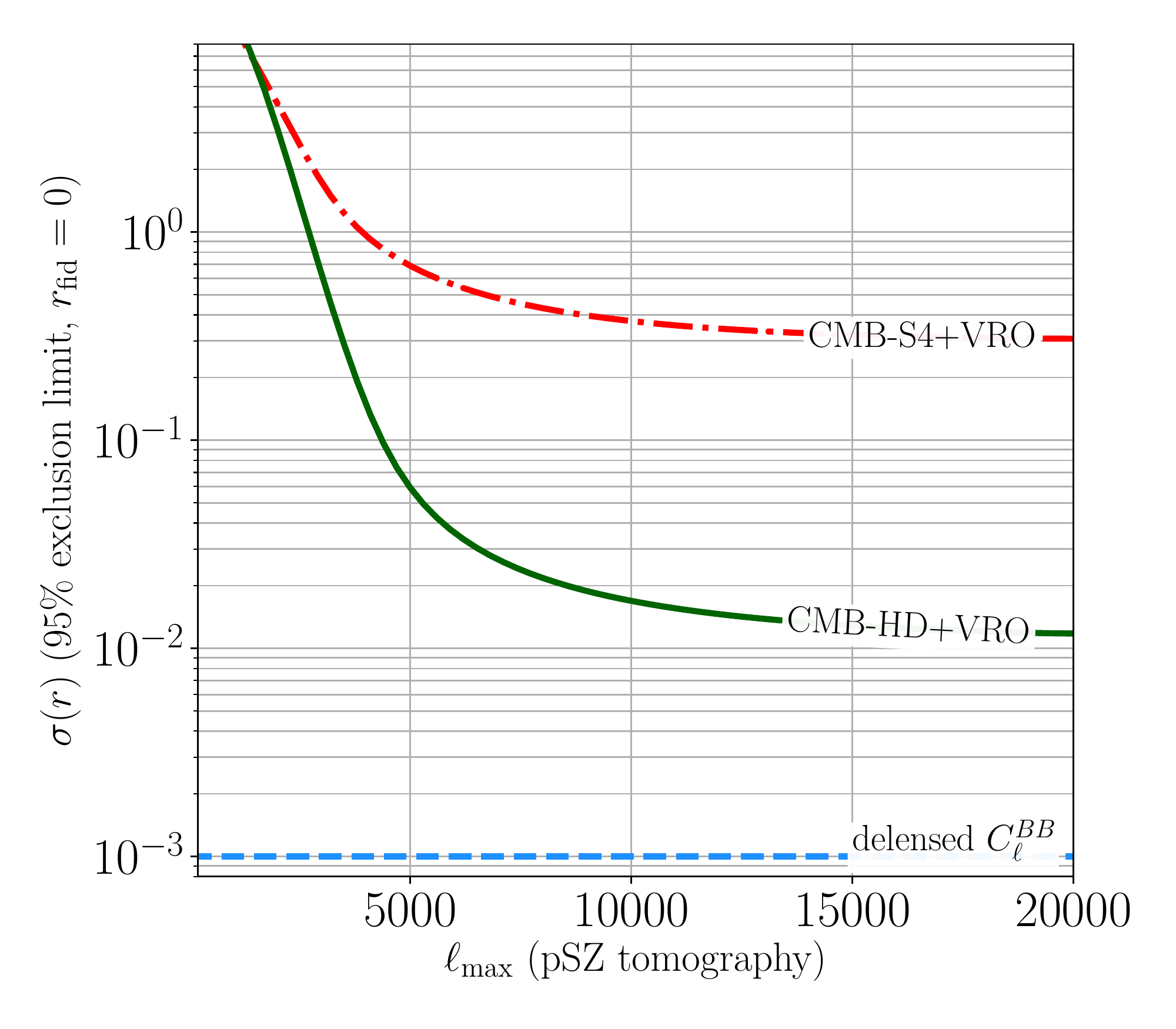}
\caption{{\bf{Inflation:}} The statistical power of the pSZ tomography is shown with forecasts for the $95\%$ exclusion limit on the tensor-to-scalar ratio $r$. We use a galaxy survey with specifications matching the upcoming Vera Rubin Observatory (VRO) together with CMB surveys with specifications matching CMB-HD (solid green line) and CMB-S4 (dot-dashed red line). The dashed blue line corresponds to the anticipated constraints from the delensed CMB B-mode searches on large-scales with CMB-S4. For the galaxy survey, we take 32 redshift bins within the range $0.1<z<6.0$. The photo-z errors satisfy $\sigma_z=0.03$ and we model the galaxy bias by $bg(z)=0.95/D(z)$ where $D(z)$ is the matter growth function. We assume the distribution of electrons trace dark matter and set the survey size to $f_{\rm sky}=0.5$. We assume the remote quadrupole can be reconstructed for modes satisfying $2 \leq L\leq 20$.  The x-axis shows the increase in the constraining power of pSZ tomography from the non-Gaussian information provided by the smallest scales accessible to CMB-HD.  {\it{Figure credit: Selim~C.~Hotinli.}}}
\label{fig:CMB_HD_r}  
\end{figure}

The search for primordial gravitational waves (PGWs) through measuring their impact on CMB polarization is one of the primary science goals of current and the upcoming CMB missions. For the current missions, the most promising route for the detection of the PGWs is through measuring their impact on the contribution to the B-mode fluctuations on scales $40 \lsim L \lsim 1000$. For these searches, confusion from polarised Galactic dust emission and weak gravitational lensing by large-scale structure are significant obstacles, with the former limiting the observable sky to a small fraction of the survey area. Dedicated studies suggest that the tensor-to-scalar ratio $r$ will be constrained below $\lesssim 0.001$ at $95\%$ significance with the near-future CMB experiments. The ability of CMB-HD to measure small-scales to high accuracy provides an excellent window of opportunity to improve these errors through a higher-quality measurement of small-scale CMB fluctuations that can be used to remove the lensing-induced $B$-mode polarisation more efficiently. Using CMB-HD to delens maps of upcoming satellite missions, such as LiteBird for example, is one of the possible routes to potentially contribute to PGW searches.

In addition to inducing large-scale B-mode fluctuations, PWGs leave a distinct imprint on the remote quadrupole field — the CMB quadrupole observed from different locations in the Universe~\citep{2012PhRvD..85l3540A}. The remote quadrupole field can be measured via observing the variations of the polarised Sunyaev Zeldovich (pSZ) effect in the CMB maps~\citep{2012PhRvD..85l3540A,Deutsch:2018umo,Deutsch:2017ybc,Deutsch:2017cja}. The pSZ effect is a small-scale CMB polarization anisotropy induced by the scattering of CMB photons off energetic free electrons in the post-reionization Universe. The cross-correlation of the pSZ effect with galaxy surveys (a technique referred to as pSZ tomography) can be used to reconstruct the remote quadrupole field and hence to measure the PWGs~\citep{2012PhRvD..85l3540A,Deutsch:2018umo}. The high precision of CMB-HD puts it in a unique position to detect this small-scale signal and constrain PWGs, providing a valuable method alternative to large-scale B-mode searches. Pushing PWG searches to a `multiple-observable’ regime, pSZ tomography with CMB-HD can provide a unique way to cross-validate existing PWG measurements (or constraints) from the primary CMB, and contribute to the characterisation of PWGs beyond exclusion limits. 

In Figure~\ref{fig:CMB_HD_r} we demonstrate the statistical power of pSZ tomography using a galaxy survey with specifications matching the upcoming Vera Rubin Observatory (VRO) with CMB surveys with specifications matching CMB-S4 (dot-dashed red line) and CMB-HD (solid green line). The ability to probe smaller scales with CMB-HD significantly improves the prospects to detect PWGs with pSZ tomography compared to CMB-S4, suggesting constraints comparable to primary CMB B-mode searches may be achievable in the future. The dashed blue line corresponds to anticipated constraints from primary CMB B-mode searches with CMB-S4. In our forecasts we followed~\citep{Deutsch:2018umo} to calculate the tensor quadrupole signal and the reconstruction noise from pSZ tomography.  We assume the remote quadrupole can be reconstructed for modes satisfying $2 \leq L\leq 20$. As the dominant contribution to the pSZ signal-to-noise comes from the largest scales, the measurement quality of modes around $L\sim2$ will play an important role in capitalising from the statistical power of pSZ tomography, and the limited survey size can lead to a more significant effect compared to $f_{\rm{sky}}$~\citep{2012PhRvD..85l3540A}. Kinematic contributions to the CMB polarization on small scales can potentially bias or add noise to the small-scale pSZ measurements, while deviations of the electron distribution from matter may reduce the statistical power available to the pSZ measurement below arcminute scales. For the currently accepted electron models we anticipate the increase in errors from the former could be around a factor of 2. Future measurements of electron profiles and distribution from ongoing FRB searches as well as kSZ and tSZ measurements, for example, can potentially improve the statistical power of pSZ tomography. Simulation-based studies of the prospects to reconstruct the remote quadrupole from the CMB can help further demonstrate the benefits of CMB-HD.

\section{Dark Energy and Neutrino Mass}
\label{sec:DE_Mnu}

CMB-HD will generate a catalog of galaxy clusters containing close to half a million objects, increasing the sample size by several fold compared to current surveys.  These galaxy clusters will be detected via the thermal Sunyaev-Zel'dovich (tSZ) effect, which occurs due to inverse Compton scattering of CMB photons off the ionized gas in galaxy clusters. This cluster catalog will be mass-limited down to $\mvir \sim 3-5 \times 10^{13}\ \msol$ with a well understood selection function even at high redshifts ($z \ge 2$) \citep{raghunathan21a, raghunathan21b}. Since both dark energy and the free-streaming length of neutrinos affect the small-scale matter power spectrum, measuring the number of clusters in multiple mass and redshift bins -- i.e.~the cluster abundance -- is an excellent probe of both of these parameters. These cluster abundance measurements in combination with primary CMB power spectra ($TT/EE/TE$) offer compelling constraints on both the sum of the neutrino masses, $\summnu$, and the dark energy equation of state, $\wde$. One of the primary challenges for cluster abundance measurements is the unbiased mass calibration of galaxy clusters. CMB-HD, with its exquisite measurements of the small-scale lensing of the CMB temperature and polarization anisotropies, will also provide sub-percent level errors on the masses of the detected galaxy clusters. 

\begin{figure}[t]
\includegraphics[width=0.49\textwidth,keepaspectratio]{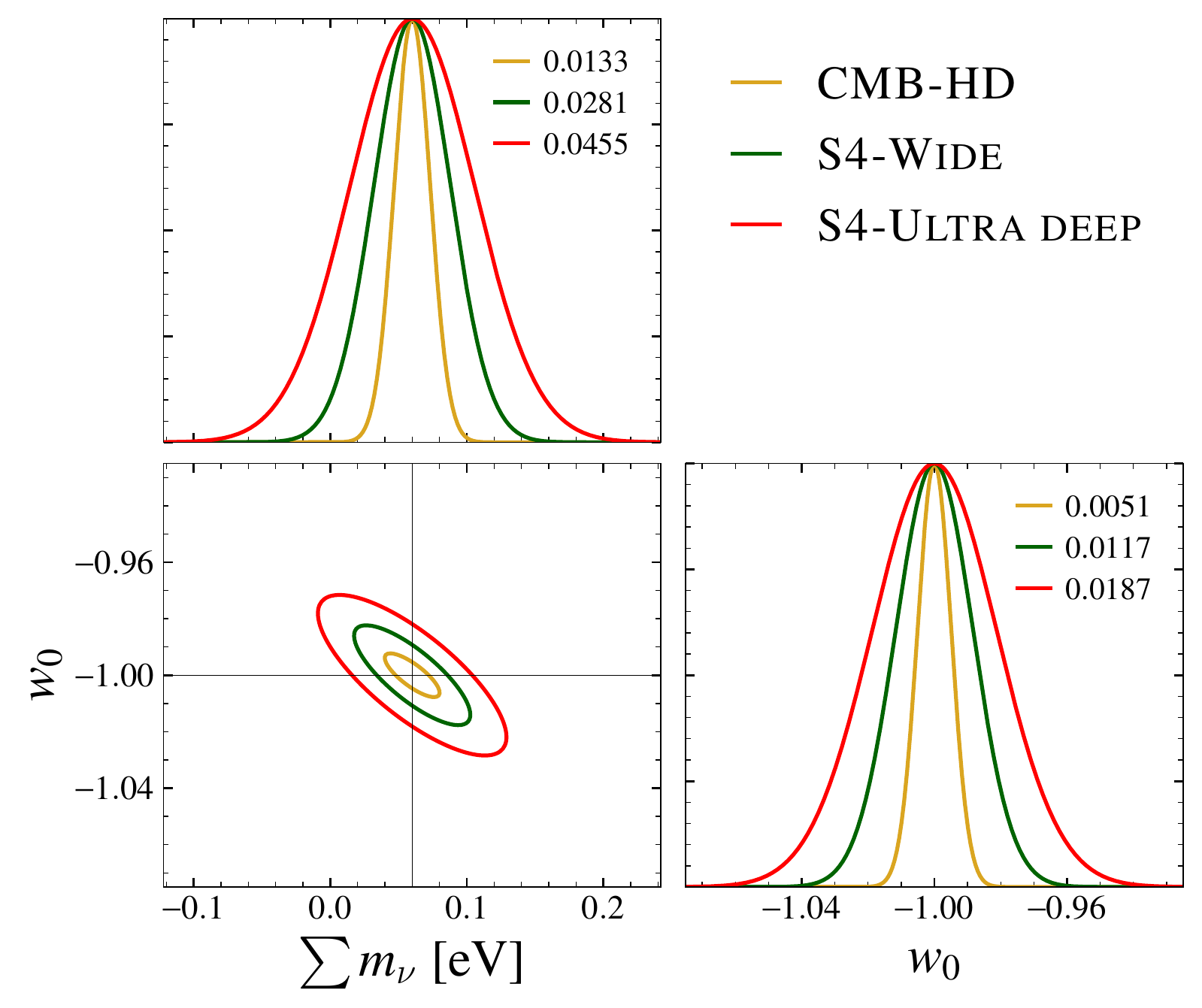}
\hspace{5mm}
\includegraphics[width=0.45\textwidth,height=7.0cm]{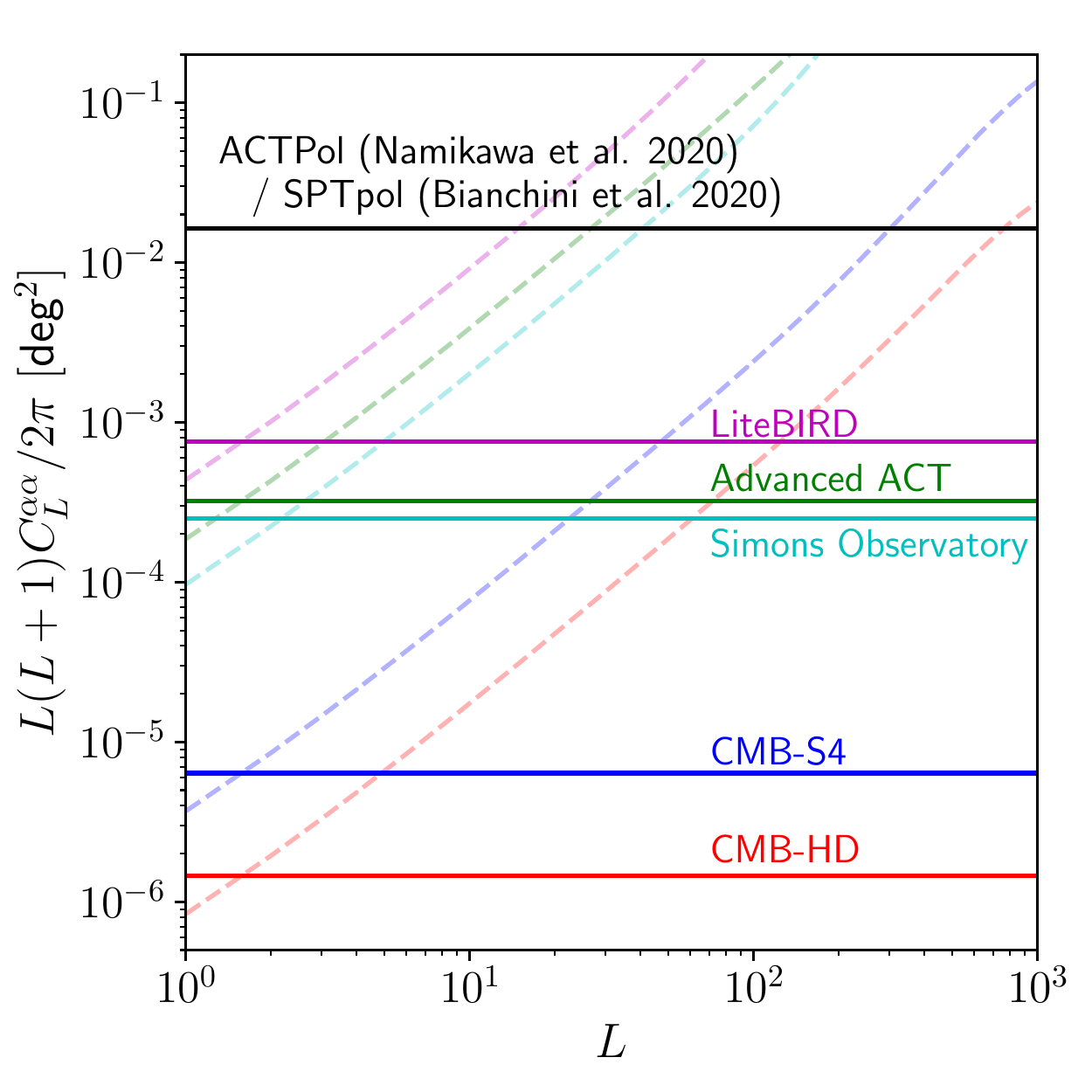}
\caption{{\it{Left:}} {\bf{Dark Energy and Neutrino Mass:}} Marginalized constraints on the dark energy equation of state parameter, $w_{0}$, and the sum of the neutrino masses, $\sum m_{\nu}$, for CMB-HD (yellow) and CMB-S4 surveys. For CMB-S4, both the Chilean wide (S4-Wide in green) and the South Pole deep (S4-Ultra deep in red) surveys are shown. The constraints are obtained by combining primary CMB power spectra ($TT/EE/TE$) with cluster abundance measurements. A {\it Planck}-like $\tau_{\rm re}$ prior of $\sigma(\taure)= 0.007$ is assumed. {\it{Figure credit: Srinivasan Raghunathan.}}
{\it{Right:}} {\bf{Beyond Standard Model:}} Expected constraints on anisotropic cosmic birefringence. The horizontal solid colored lines show the forecasted $68\%$ CL bounds on the scale-invariant power spectrum, using the statistical uncertainty on the birefringence power spectrum for each experiment shown by the dot-dashed colored lines. The black solid line shows the current $68\%$ CL upper bound. Cosmic birefringence can constrain very light axion-like particles, the axion string network, axion dark matter, general Lorentz-violating physics in the context of Standard Model extensions, and primordial magnetic fields. {\it{Figure credit: Toshiya Namikawa.}}}
    \label{Fig:DE_Mnu_biref}
\end{figure}

Marginalized joint constrains on the dark energy equation of state, $\wde$, and the sum of the neutrino masses, $\summnu$, are presented in the left panel of Figure~\ref{Fig:DE_Mnu_biref}. The constraints are obtained by combining primary CMB power spectra ($TT/EE/TE$) measurements with cluster abundance measurements using an internal CMB-lensing mass calibration. CMB-HD, shown in yellow, can obtain errors of $\sigma(\wde)= 0.005$ and $\sigma(\summnu)=13$~meV~\citep{raghunathan21a, raghunathan21b}. Given the current lower bound of $\summnu \ge 58$~meV, this implies CMB-HD will be able to detect the sum of neutrino masses at about $5\sigma$ significance or higher. Other cosmological parameters and the ones governing the observable-mass scaling relation of galaxy clusters have been marginalized over and are not shown in the figure. A {\it Planck}-like prior on the optical depth to reionization of $\sigma(\taure) = 0.007$ has been assumed. Assuming $\sigma(\taure) = 0.002$, as expected from the proposed LiteBIRD or CORE \citep{hazra18, divalentino18, Brinckmann:2018owf} missions, can improve the constraints further by a factor of 1.5. For reference, expected constraints from two surveys (S4-Wide survey from Chile in green and S4-Ultra deep survey from the South Pole in red) of the proposed CMB-S4 experiment are also shown. 

CMB lensing power spectrum measurements are also a sensitive probe of the sum of the neutrino masses, and offer an independent pathway to $\summnu$ constraints.  While these measurements have a degeneracy between $\summnu$ and other cosmological parameters that affect structure formation, like the matter density, $\Omega_{m}$, or $\wde$~\citep{madhavacheril17}, such degeneracies can be broken by including external datasets, such as from baryon acoustic oscillations.

\section{Other Beyond Standard Model}

\subsection{Cosmic Birefringence}
\label{sec:Biref}

CMB polarization data can be used to test for new physics by searching for a rotation of linear polarization as the CMB photons propagate to us from the CMB last scattering, usually referred to as cosmic birefringence. The source of the cosmic birefringence includes axion-like particles with a very small mass $m_a\lesssim 10^{-28}$\,eV coupled with photons through the Chern-Simons term \cite{Carroll:1998,Li:2008,Pospelov:2009,Capparelli:2019:CB}, the axion string network \cite{Agrawal:2019:biref}, axion dark matter \cite{Liu:2016dcg}, general Lorentz-violating physics in the context of Standard Model extensions \cite{Leon:2017}, and primordial magnetic fields through Faraday rotation of the CMB polarization  \cite{Kosowsky:1996,Harari:1997,Kosowsky:2004:FR}.
For example, parity violating couplings between photons and axion-like pseudoscalar particles lead to polarization and magnetization contributions to the electromagnetic field, and if the axion field varies spatially or temporally, then the plane of polarizaton of the CMB photons is rotated (see Sec. 9.9 of axion review~\cite{Marsh2016}).
It is seen that for axions lighter than the Hubble scale at decoupling, i.e. $10^{-28}$\,eV, the contribution to cosmic birefringence is significant \cite{Marsh2016}.
Measurement of cosmic birefringence can also constrain the coupling $g_{\alpha\gamma}$ of axion-like dark matter to photons. These constraints scale weakly with the fraction $F=\Omega_a/\Omega_c$ of the energy density of axion-like dark matter to the total dark matter, such that $g_{\alpha\gamma}\propto F^{-1/2}$~\cite{Sigl:2018fba}.

If the source of the cosmic birefringence is spatially varying on the sky, the polarization rotation will be anisotropic \cite{Carroll:1998,Lue:1999,Li:2008,Caldwell:2011,Lee:2015,Leon:2017}.
The isotropic and anisotropic birefringence measurements are complementary probes in order to constrain models which source cosmic birefringence. For example, quintessence models predict both isotropic and anisotropic cosmic birefringence \cite{Caldwell:2011,Capparelli:2019:CB}. In addition, the cosmic birefringence induced by some massless scalar fields is not necessarily isotropic, and a measurement of the anisotropic birefringence is crucial to constraining such scenarios~\cite{Gluscevic:2009,2012PhRvD..86j3529G}. Measurements of, or tight constraints on, the relevant pseudo-scalar fields and other phenomena can hence provide valuable insights into fundamental physics. 

Both isotropic and anisotropic cosmic birefringence have been constrained by several CMB experiments. 
The isotropic cosmic birefringence measurements have utilized $B$-modes, part of which are converted from $E$-modes by the cosmic birefringence effect \cite{Feng:2006:CPT,Gruppuso:2012,Li:2014,Mei:2014iaa}. This $E$-to-$B$ leakage introduces non-zero odd-parity $TB$ and $EB$ power spectra which are zero in the standard cosmological model. The systematic errors of the uniform polarization angle calibration in several existing CMB data are, however, larger than the statistical error \cite{Pagano:2009:biref,Miller:2009:biref,Keating:2013,Hinshaw:2013,P16:rot}. To circumvent the situation, we can also use Galactic foreground contributions in the observed odd-parity spectra to partially break degeneracies between the polarization angle error and cosmic birefringence signals \cite{Minami:2019:rot}, and this technique was demonstrated using Planck data \cite{Minami:2020:biref}. 
An alternative way is to use a mode coupling introduced by gravitational lensing in the presence of the isotropic cosmic birefringence. The expected sensitivity to the isotropic birefringence using the mode coupling is comparable to that using the conventional $TB$ and $EB$ spectra. We can use these two methods as a cross-check for the isotropic birefringence measurement. 

Anisotropic cosmic birefringence is efficiently constrained by using the fact that the $EB$ correlation varies with direction. This mixes together $E$ and $B$ modes of different scales, leading to non-zero expectation values in the off-diagonal ($\ell,m\ne\ell',-m'$) elements of the CMB covariance. Cross-correlating different angular scales of $E$ and $B$ modes, we can reconstruct the anisotropies of the cosmic birefringence signals \cite{Kamionkowski:2009:derot}.\footnote{Other pairs of CMB anisotropies such as temperature and $B$-modes are also correlated, but such correlations generally give lower signal-to-noise ratios for reconstructing birefringence \cite{Yadav:2012a}.} The power spectrum of the reconstructed cosmic birefringence anisotropies is, however, a four-point correlation of the CMB anisotropies and several biases must be subtracted in future high-precision experiments \cite{Namikawa:2016:rotsim}. Since most physically well-motivated models of anisotropic birefringence predict a scale-invariant spectrum, multiple works have presented constraints on the amplitude of the scale-invariant spectrum of the cosmic birefringence using reconstruction methods; these have made use of the WMAP~$TB$ correlation \cite{Gluscevic:2009}, or the polarization data of the POLARBEAR \cite{PB15:rot}, BICEP2/Keck Array \cite{BKIX}, and Planck \cite{Contreras:2017,Gruppuso:2020:biref} experiments. The current best constraint comes from ACTPol \cite{Namikawa:2020:ACT-biref} and SPTpol \cite{Bianchini:2020:SPT-biref}. We note that, although the reconstruction technique will be the best way to constrain anisotropic birefringence, several other publications \cite{Gubitosi:2011:biref,Li:2013,Alighieri:2014yoa,Li:2014,Mei:2014iaa,Liu:2016dcg,Zhai:2019god} also place constraints on anisotropic birefringence by analyzing CMB polarization power spectra.

CMB-HD will improve the constraints on the scale-invariant anisotropic birefringence by four orders of magnitude better than the current best constraints. The right panel of Figure~\ref{Fig:DE_Mnu_biref} shows the expected constraints on the amplitude of the scale-invariant power spectrum from future CMB experiments (see also \cite{Pogosian:2019:axion,Mandal:2022tqu}).

\vspace{-2mm}
\section{Astrophysics}

\subsection{Evolution of Gas}

A fundamental question in astrophysics is: how did galaxies form? A critical advance in answering this question would be measuring the density, pressure, temperature, and velocity of the gas in and around dark matter halos out to $z\sim 2$ and with masses below $10^{12}$ M$_\odot$.  This is because the gas in these halos reflects the impact of feedback processes and mergers, and it serves as a reservoir enabling star formation.  However, to date we have not had such measurements over a statistically significant sample. 

CMB-HD would open this new window by measuring the tSZ and kSZ effects with high-resolution and low-noise over half the sky (for a recent SZ review see~\cite{Mroczkowski2019}).  The tSZ signal is a measure of the thermal pressure of ionized gas in and around dark matter halos~\cite{Sunyaev1970,Sunyaev1972}.  The kSZ signal measures the gas momentum density~\cite{Sunyaev1980}. The combination of both tSZ and kSZ measurements, with low-noise and high-resolution across CMB-HD frequencies,
would allow CMB-HD to separately measure the density, pressure, temperature, and velocity profiles of the gas, as a function of halo mass and redshift~\cite{Knox2004,Sehgal2005,Battaglia2017}.  This would probe thermal, non-thermal, and non-equilibrium processes associated with cosmic accretion, merger dynamics, and energy feedback from stars and supermassive black holes, and their impact on the gas~\cite{Nagai2011,Nelson2014b,Lau2015,Avestruz2015,Basu2016}.

Measuring the tSZ effect is also an effective way to find new galaxy clusters and groups, as has been well-demonstrated over the past decade~\cite{Marriage2011,Vanderlinde2010,Planck2014}. 
Millimeter-wave SZ measurements have an advantage over X-ray measurements in probing gas physics because the SZ signals are proportional to the gas density (not density squared) and the brightness of the signals are redshift independent.  This makes SZ measurements a powerful probe of the gas in the outskirts of galaxy clusters, and in low-mass and high-redshift halos.  CMB-HD will push halo-finding to lower masses and higher redshifts, allowing direct imaging of systems where X-ray observations would require prohibitively long integration times. By stacking the tSZ-detected halos, CMB-HD enables probing the gas physics in and around halos out to $z\sim 2$ and with masses below $10^{12}$ M$_\odot$.  The circumgalactic reservoirs of $10^{12}$ M$_\odot$ (Milky-Way-mass) halos are predicted by multiple simulations, such as EAGLE and Illustris-TNG, to be intimately linked to the appearance of, and activity within, the galaxy~\cite{schaye2015,nelson2018a,davies2019,pillepich2018}.  These and other simulations find that galactic star-formation rates, colors, and morphologies are inextricably linked not only to the mass in the circumgalactic medium, but also to the location of baryons ejected beyond $R_{200,c}$, which can be uniquely constrained by CMB-HD. Thus the science gain of such measurements is a more complete understanding of galaxy cluster astrophysics, the physics of the intergalactic and circumgalactic medium, and galaxy evolution.

\subsection{Planetary Studies}

The complete inventory of planets, dwarf planets, and asteroids in our own Solar System remains an open question.  CMB-HD can open a new discovery space in our outer Solar System by detecting undiscovered Solar System bodies via their thermal flux and parallactic motion~\cite{Cowan2016}. Objects close to the Sun are normally detected via optical observations, which are sensitive to the bodies' reflected light from the Sun.  However, objects also have internal heat that is emitted at millimeter wavelengths.  Since the flux from reflected light falls faster with distance than directly sourced emission, CMB-HD has an advantage over optical surveys in finding objects in the far Solar System~\cite{Baxter2018a}. In particular, CMB-HD could detect dwarf-size planets hundreds of AU from the Sun, and Earth-sized planets thousands of AU from the Sun.  In combination with  optical measurements, CMB-HD would allow large population studies of the sizes and albedos of these objects~\cite{Gerdes:2017}. In addition, whether our Solar System, and exo-solar systems in general, possess an Oort cloud is still unknown.  The low-noise and high-resolution of CMB-HD would enable the detection of exo-Oort clouds around other stars, opening a new window on planetary studies.

\subsection{Transients}

CMB-HD will plan to cover 50\% of the sky with a daily cadence.  The sensitivity of CMB-HD will result in an $8\sigma$ daily flux limit of 4 mJy at 150 GHz.  At this flux limit, we expect an average of 0.1 false detections from random noise fluctuations over the full survey of 7.5 years. Note that for moving or variable sources, source confusion from other static sources is not an issue.  The increased sensitivity of CMB-HD also allows one to probe more volume to detect the same events compared to precursor surveys by detecting the same sources at larger cosmological distances. Being 15 times deeper in flux compared to CMB-S4 gains 60 times more volume to probe sources of the same intrinsic luminosity.  

Thus CMB-HD will have excellent sensitivity to bright time-variable and transient sources in the sky.  For example, CMB-HD will detect of order 100 on-axis long gamma-ray bursts (LGRBs)~\cite{Metzger2015}.  These LGRBs often are bright in the millimeter weeks before they peak in the radio, and the additional frequency coverage can help characterize their shock behavior~\cite{Laskar2018}.  Jet-dominated active galactic nuclei called blazars are bright in the millimeter and vary significantly at about 100 GHz on week timescales~\cite{Chen2013,Madejski2016,Blandford2018}.  Novae, which are repeating thermonuclear explosions of accreting white dwarfs, are also millimeter-bright, with expected rates of about ten per year in our Galaxy.  CMB-HD will provide unique insight into the geometry of blazars and novae by providing polarization flux and variability information for these systems~\cite{Abdo2010,Young2010}. In addition, some neutron star mergers may be seen and localized first in the millimeter band~\cite{Fong2015}.  The CMB-HD survey would thus provide useful follow-up of LIGO/Virgo triggers, and could provide blind detections of these events~\cite{Holder2019}. Further details on the types of transient sources CMB-HD is sensitive to can be found in~\cite{Metzger:2015eua,2019BAAS...51c.331H,Eftekhari:2021xgu,chichura2022asteroid}.  The intent of the CMB-HD project is to provide to the astronomy community weekly maps of the CMB-HD survey footprint, filtered to keep only small scales, and with a reference map subtracted to make variability apparent.

\subsection{Catalog of Galaxies}

CMB-HD's contribution to the discovery of particularly rare, dusty galaxies may also be significant given its coverage of a large fraction of the sky at millimeter wavelengths.  Dusty Star-Forming Galaxies (DSFGs: see~\cite{Blain2002,Casey2014} for reviews) with star formation rates in excess of 100\,$M_\odot$\,yr$^{-1}$ constitute some of the most luminous and massive galaxies in the Universe.  As prodigious star-forming galaxies they produce incredibly massive reservoirs of dust, even at early times in the Universe's history ($z>5-6$;~\cite{Zavala2018,Strandet2017,Marrone2018}), such that their quick formation poses unique challenges for cosmological models of massive galaxy formation, as well as our understanding of dust production mechanisms from AGB stars, supernovae, and ISM grain growth.  While DSFGs are a well-understood population of massive star-forming galaxies that dominate cosmic star-formation at its peak $2<z<3$~\cite{Gruppioni2013,Chapman2005,Danielson2017}, the relative abundance of rare DSFGs at earlier times is not as well quantified~\cite{Zavala2018,Williams2019,Casey:2021faq}.  And though several DSFGs have been identified at such early times~\cite{Cooray2014,Zavala2018,Marrone2018,Casey2019}, the detection and characterization of such rare, extreme sources necessitates an ambitious millimeter survey covering a substantial fraction of the sky at depths needed to pioneer their detection.

While the majority of DSFGs have been identified at wavelengths $\sim$250$\mu$m--1\,mm, CMB-HD's depth at longer wavelengths (2--3\,mm) translates to a more effective strategy in identifying the highest redshift DSFGs by filtering out the foreground~\cite{Casey2019b}.  The primary limitation of CMB-HD will be its angular resolution, limited to $\sim$0.25\,arcmin in beamsize. The angular resolution limits analysis of galaxies that are relatively faint, but the unique contribution of CMB-HD is its ability to directly detect and characterize brighter, rarer DSFGs with S$_{2mm}>0.5$\,mJy ($10\sigma$ limit) with source density 2$\times$10$^2$\,deg$^{-2}$, a quarter of which will likely be at $z>4$~\cite{Zavala2021,Casey:2021faq}.  Constraining the source density of such rare, early galaxies is critical to understanding their early formation; swift follow-up with interferometric observations can then lead to greater insight regarding the physical evolution of their interstellar media, shedding light on what makes these systems so unique in the first instance.

\section{Technical Overview}

\subsection{Measurement Requirements}
\label{sec:measureReq}

The technical requirements are driven by the dark matter and new light species science targets, since those targets set the most stringent requirements.  All the other science targets benefit from and prefer these same technical requirements, since there are no science targets that pull the requirements in opposing directions.  This results in a fortunate confluence between the science targets presented above and the technical requirements that can achieve them. \\

\noindent {\bf{Sky Area:}} Both the dark matter and new light species science targets prefer the widest sky area achievable from the ground, given fixed observing time~\cite{Sehgal2019a,CMBS4SB}.  For the dark matter science case, the inclusion of the kSZ foregrounds is what makes wider sky areas preferable over smaller ones~\cite{Sehgal2019a}.  In addition, the non-Gaussianity inflation science target, searches for planets and dwarf planets, and probing the transient sky all benefit from the largest sky areas possible.  In practice, this is about {\it{\underline{50\% of the sky}}}, achievable from the demonstrated site of the Atacama Desert in Chile. \\

\noindent {\bf{Resolution:}} The resolution is set by the dark matter science target of measuring the matter power spectrum on comoving scales of $k\sim10 h$Mpc$^{-1}$ (these scales collapsed to form masses below $10^9 M_\odot$ today) (see Section~\ref{sec:grav} for details).  Since CMB lensing is most sensitive to structures at $z\sim 2$ (comoving distance away of 5000 Mpc), we need to measure a maximum angular scale of $\ell \sim k X \sim 35,000$. {\it{\underline{This gives a required resolution of about 15 arcseconds at 150 GHz}}} ($\ell = \pi$/radians), translating to a 30-meter telescope.  In addition, foreground cleaning will be essential, and the most dominant foreground will be extragalactic star-forming galaxies, i.e.~the Cosmic Infrared Background (CIB).  Figure~\ref{fig:foregrounds} shows that removing sources above a flux cut of 0.03 mJy at 150 GHz lowers the CIB power to well below the instrument noise.  A resolution of 15 arcsecond is also needed to measure the profiles of the gas in halos and to separate extragalactic radio and star-forming galaxies from the gas signal.  To obtain a census of objects in the outer Solar System, we note that the parallactic motion of objects 10,000 AU away from the Sun is about 40 arcseconds in a year, also requiring this minimum resolution to detect the motion across a few resolution elements. \\

\noindent {\bf{Sensitivity:}} The sensitivity is driven by the dark matter science target. In order to detect with $5\sigma$ significance a deviation of the matter power spectrum from the CDM-only prediction at a level that matches claimed observations of suppressed structure, one {\it{\underline{requires 0.5~$\mu$K-arcmin instrument noise}}\\{\underline{in temperature (0.7~$\mu$K-arcmin in polarization) in a combined 90/150 GHz channel.}}} This assumes the residual foreground levels shown in Figure~\ref{fig:foregrounds}~\cite{Han:2021vtm}.
Conveniently, this sensitivity level also allows one to cross critical thresholds, achieving $\sigma({N_{\rm{eff}}}) = 0.014$, $\sigma(f_\mathrm{NL}^{\rm local}) = 0.26$, and $\sigma(r)= 0.005$.  \\

\begin{table}[t]
\centering
\caption{Summary of Measurement Requirements for CMB-HD Survey Over Half the Sky}
\begin{tabular}[t]{|l|ccccccc|}
\hline
Frequency (GHz)~~   & 30 & 40  & 90  & 150  & 220  & 280  & 350  \\
Resolution (arcmin)~~  & 1.25~  & 0.94~  & 0.42~  & 0.25~  & 0.17~  & 0.13~  & 0.11~  \\
White noise level ($\mu$K-arcmin)\textsuperscript{a}~~ & 6.5  & 3.4  & 0.7  & 0.8  & 2.0  & 2.7  & 100.0  \\
\hline
\end{tabular}
\begin{tablenotes}
\item \textsuperscript{a} 
Sensitivity is for temperature maps.  For polarization maps, the noise is $\sqrt{2}$ higher.
\end{tablenotes}
\label{tab:cmbhdSpecs}
\end{table}

\noindent {\bf{Largest Angular Scale:}} By the time CMB-HD has first light, data from the Simons Observatory (SO)~\cite{SO2019}, which will have first light in 2024, will already exist and be public.  SO will measure temperature and E-mode maps over half the sky to the sample variance limit in the multipole range of 30 to 3000 for temperature and 30 to 2000 for E-modes.  SO will also measure these scales at six different frequencies spanning 30 to 280 GHz.  Thus there is no need for CMB-HD to reimage these modes.  Therefore the largest angular scale CMB-HD needs to measure is driven by measuring B-modes.  For all the science goals discussed above, with the exception of the measurements of cosmic birefringence and inflationary magnetic fields, the largest scale CMB-HD needs to image is about 10 arcminutes ($\ell \sim 1000)$.  However, to achieve the birefringence and inflationary magnetic field measurements (see Sections~\ref{sec:PMFs} and~\ref{sec:Biref}) requires CMB-HD to measure polarization anisotropies down to $\ell \sim 100$. {\it{\underline{This requires that the minimum multipole in temperature maps be}}\\ {\underline{$\ell \sim 1000$, and the minimum multipole in polarization maps be $\ell \sim 100$}}}. \\

\noindent {\bf{Frequency Coverage:}} The frequency coverage is driven by needing most of the sensitivity in the frequency window that is most free from extragalactic foregrounds, namely 90 to 150 GHz.  Modern detectors can observe at two frequencies simultaneously~\cite{Henderson2016,Benson2014,OBrient2013}, so we assume we can split closely spaced frequency bands, further helping to remove frequency-dependent foregrounds.  We also require frequency coverage at 30/40 GHz to remove emission from radio galaxies and at 220/280 GHz to remove emission from dusty galaxies and to cover the null frequency of the tSZ signal (at 220 GHz).  Foreground optimization studies done for SO have found optimal ratios of noise levels given their six frequency channels, which if extrapolated to {\it{\underline{CMB-HD would require}}\\ {\underline{noise levels in temperature maps of 6.5/3.4, 0.73/0.79, and 2/4.6 $\mu$K-arcmin for the 30/40, 90/150,}}\\ {\underline{and 220/280 GHz channels respectively.}}} In addition, CMB-HD requires deeper sensitivity at 280 GHz in order to better remove one of the main foregrounds (the CIB).  To clean the CIB to the level shown in Figure~\ref{fig:foregrounds} 
{\it{\underline{requires an additional 280 GHz channel with 3.25~$\mu$K-arcmin noise, yielding a}}\\ {\underline{combined noise level of 2.65~$\mu$K-arcmin at 280 GHz}}} (see Section~\ref{sec:fg} for more detail).  This 280 GHz channel can be split into a 280/350 GHz channel at minimal cost in order to gain a 350 GHz channel with a noise level of about 100~$\mu$K-arcmin noise.

\subsection{Foreground Removal}
\label{sec:fg}

\begin{SCfigure}[1.4][t]
\centering
\includegraphics[width=0.55\textwidth,height=7.3cm]{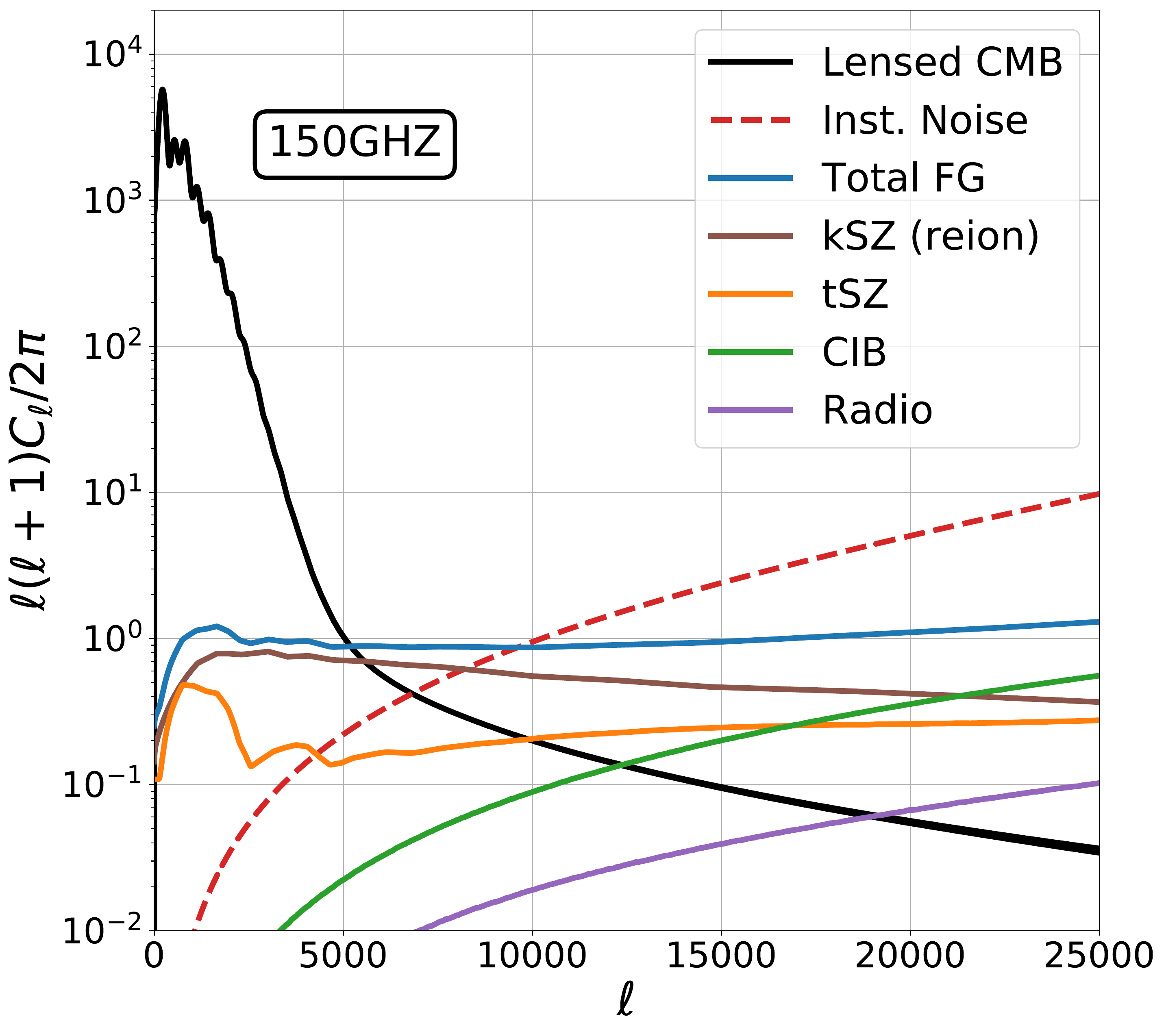}
\caption{Shown are the CMB temperature power spectrum (black solid) and relevant foregrounds at 150~GHz.  The foregrounds shown are the kSZ effect from the epoch of reionization (brown), the tSZ effect (orange), the CIB (green), and radio sources (purple) (after removing CIB and radio sources above a flux of 0.03 mJy).  The CMB-HD instrument noise at 150~GHz is 0.8~$\mu$K-arcmin (dashed red), and combining 150 and 90 GHz channels brings the effective noise level to 0.5~$\mu$K-arcmin.}
\label{fig:foregrounds}  
\vspace{-5mm}
\end{SCfigure}

One of the largest challenges to achieve the key science aimed for the by CMB-HD experiment is removal of astrophysical foregrounds.  Extraglactic foregrounds consist of the thermal and kinetic Sunyaev-Zel’dovich effects (tSZ and kSZ), the Cosmic Infrared Background (CIB), and radio galaxies (Radio).  Galactic foregrounds consist primarily of dust and synchrotron emission from the Milky Way.  

Galactic foregrounds dominate at large scales; since CMB-HD is targeting a minimum multipole of $\ell=1000$ for temperature and E-mode polarization maps, the Galactic foregrounds are largely cut out of the maps with the remainder being subdominant.  This is not the case for B-mode polarization maps however, for which CMB-HD is targeting a minimum multipole of $\ell=100$; in this case, the seven frequency channels will be used to remove the frequency-dependent Galactic signal.

Extragalactic foregrounds primarily impact temperature maps, and in many cases are both an important signal for extracting fundamental physics as well as a contaminant to other signals.  The CIB consists of dusty, star-forming galaxies emitting primarily at sub-millimeter wavelengths, and radio galaxies are largely active galactic nuclei or other sources emitting primarily at radio wavelengths.  Both of these types of sources have significant millimeter emission, and given the resolution of CMB-HD, they have a profile that matches the instrumental beam.  Table~\ref{tab:fluxlimits} shows the expected single-frequency $5\sigma$ flux limits CMB-HD will achieve between 30 and 280 GHz, assuming white noise levels only.

Given the flux limits in Table~\ref{tab:fluxlimits}, we assume that radio sources can be detected and removed that are above 0.04 mJy at 90 GHz, and correspondingly 0.03 mJy at 150 GHz, assuming a spectral in- dex of -0.8 and extrapolating the 90 GHz flux to 150 GHz.  This detection threshold assumes the absence of source confusion from blended sources; for reference, we expect the number density of radio sources to be less than 0.07 radio source per 0.42 arcminute beam (90 GHz resolution) and 0.03 radio source per 0.25 arcmin beam (150 GHz resolution.)  For CIB sources, we exploit the fact that CMB-HD will have a 280 GHz channel with 2.7 $\mu$K-arcmin white noise. We find that CIB sources above 0.15 mJy at 280 GHz can be detected at the $5\sigma$ level in the 280 GHz channel; this detection level was determined by applying a matched-filter on maps that included confusion from other CIB sources, as well as the kSZ, tSZ, and CMB. Assuming a spectral index of 2.6 for CIB sources, this results in the identification of sources above 0.03 mJy at 150 GHz.  Once detected, we assume these sources can be removed using the measured flux of each source as well as the shape of the instrumental beam.  We also assume the removal of all $3\sigma$ detected tSZ clusters.  The largest challenge here will be that the cluster tSZ signal does not follow the profile of the instrument beam.  However, given that CMB-HD has a finer resolution than the extent of most clusters, it is well positioned to characterize mean cluster profiles.  Throughout the forecasts presented above, we do not assume the kSZ signal from reionization will be removed.  There are a number of pathways being developed to remove the late-time kSZ effect, including exploiting cross correlations with galaxy surveys as well as machine learning techniques.

Figure~\ref{fig:foregrounds} shows the expected residual levels of extragalactic foregrounds in temperature maps at 150 GHz compared to the instrument noise (dashed red) and the lensed CMB power spectrum (black).
The subtraction of Galactic and extragalactic foregrounds will of course not be exact, and fully characterizing the residual levels of contamination, exploring novel foreground subtraction avenues, and exploiting knowledge from precursor surveys such as ACT, SPT, SO, SPO, and CCAT-p will all be pursued.

\begin{table}[t]
\centering
\caption{CMB-HD Single-frequency $5\sigma$ Flux Limits Assuming White Noise}
\begin{tabular}[t]{|l|cccccc|}
\hline
Frequency (GHz)~~   & 30 & 40  & 90  & 150  & 220  & 280   \\
Flux limit (mJy) & 0.14 & 0.1 & 0.04 & 0.05 & 0.1 & 0.1 \\
\hline
\end{tabular}
\label{tab:fluxlimits}
\end{table}

\subsection{Instrument Requirements}
\label{sec:instReq}

Given the technical requirements above needed to achieve the science targets, the following instrument specifications below are required. 

\vspace{3mm}
\noindent {\bf{Site:}} The required sky area and sensitivity make {\it{\underline{Cerro Toco in the Atacama Desert the best site for}\\{\underline{CMB-HD}}}}.  An instrument at this site can observe the required 50\% of the sky; in contrast, an instrument at the South Pole can access less than half of this sky area.  The sensitivity requirement of 0.5~$\mu$K-arcmin also requires locating CMB-HD at a high, dry site with low precipitable water vapor to minimize the total number of detectors needed.  No site within the U.S. has a suitable atmosphere.  While a higher site than Cerro Toco, such as Cerro Chajnantor in the Atacama Desert, might reduce the detector count further, that may not outweigh the increased cost of the higher site.\\

\noindent {\bf{Detectors:}} To reach the required sensitivity levels of 6.5/3.4, 0.73/0.79, and 2/4.6 $\mu$K-arcmin for the 30/40, 90/150, and 220/280 GHz channels respectively, requires scaling down the SO goal noise levels by a factor of 8~\cite{SO2019}.  SO, which is at the same site as CMB-HD, requires 30,000 detectors to achieve its sensitivity levels in 5 years of observation~\cite{SO2019}.  Since the noise level scales as sqrt$(N_{\rm{det}})$ and sqrt($t_{\rm{obs}}$), {\it{\underline{CMB-HD requires 1.3 million detectors and 7.5 years of observation to}\\{\underline{reach a factor of 8 lower noise than SO across all frequencies. An extra 280/350 GHz channel with}\\{\underline{3.25 $\mu$K-arcmin noise increases the detector count to 1.6~million.}}}}} This assumes a 20\% observing efficiency, as also assumed by SO. We note, however, that this observing efficiency is likely pessimistic. ACT and POLARBEAR, precursor experiments in Chile, have usually reached 17-30\% observing efficiency, and several improvements have been identified that could increase observation efficiency. For example, some solar power could reduce downtime by about 10\%. This number also assumes a detector yield of 80\% as achieved previously; however, given the higher number of wafers and more stringent screening, it may be feasible to get a yield of 90\%.\\

\noindent {\bf{Telescope Dish Size:}} To achieve the required resolution of 15 arcseconds at about 100 GHz requires {\it{\underline{a telescope dish size of about 30-meters}}}.  The foreground cleaning discussed above also necessitates a 30-meter dish for the frequencies above 100 GHz, out to at least 350 GHz. \\

\subsection{Instrument Design}
\subsubsection{Telescope}

\begin{figure}[t]
\centering
\includegraphics[width=0.95\textwidth]{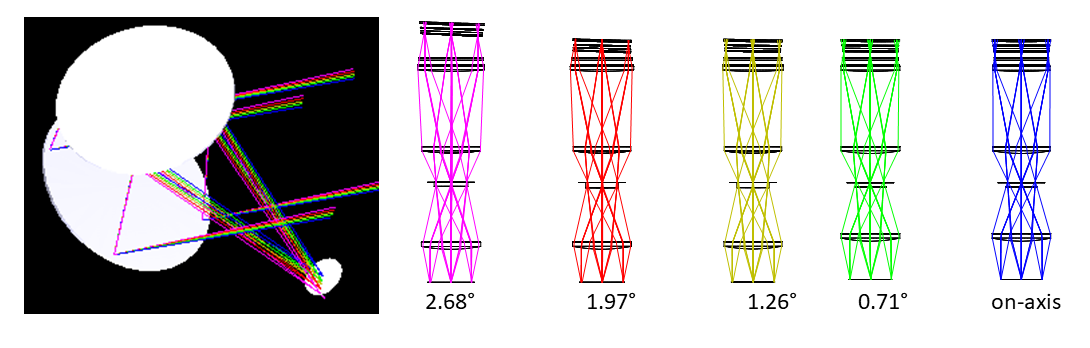}
\caption{A possible design for the telescope would be a crossed Dragone design similar to that chosen for SO and CCAT-prime.  Although challenging to build, a 30-meter version of this particular design offers a diffraction limited field of view out to $r=0.8$~degrees at 150~GHz at the secondary focus, and, with cold reimaging optics such as those shown on the right, diffraction limited beams can be achieved out past a radius of 2.68 degrees at 150~GHz and even further at lower frequencies.  These particular designs are limited by the largest silicon optics currently available (45~cm) and could be grouped together in multiple cryostats in order to ensure less down time and greater flexibility.  {\it{Figure credit: Simon Dicker}}.}
\label{fig:optics}  
\end{figure}

For systematic control, the baseline design of CMB-HD is an off-axis telescope with a primary aperature of 32 meters.  We use a crossed Dragone design because it has a larger field-of-view (fov) than Gregorian or Cassegrain telescopes.  Although the crossed Dragone design has a far larger secondary mirror (26 meters), as shown in Figure~\ref{fig:optics}, with the use of correcting cold optics extremely large fovs are possible. Simple calculations indicate that with efficient focal plane use ($>50$\%) it could be possible to fit all the detectors in one telescope.  However, given the number of receivers required, the baseline design consists of two telescopes. \\

\begin{figure}[h]
\centering
\includegraphics[width=0.8\textwidth]{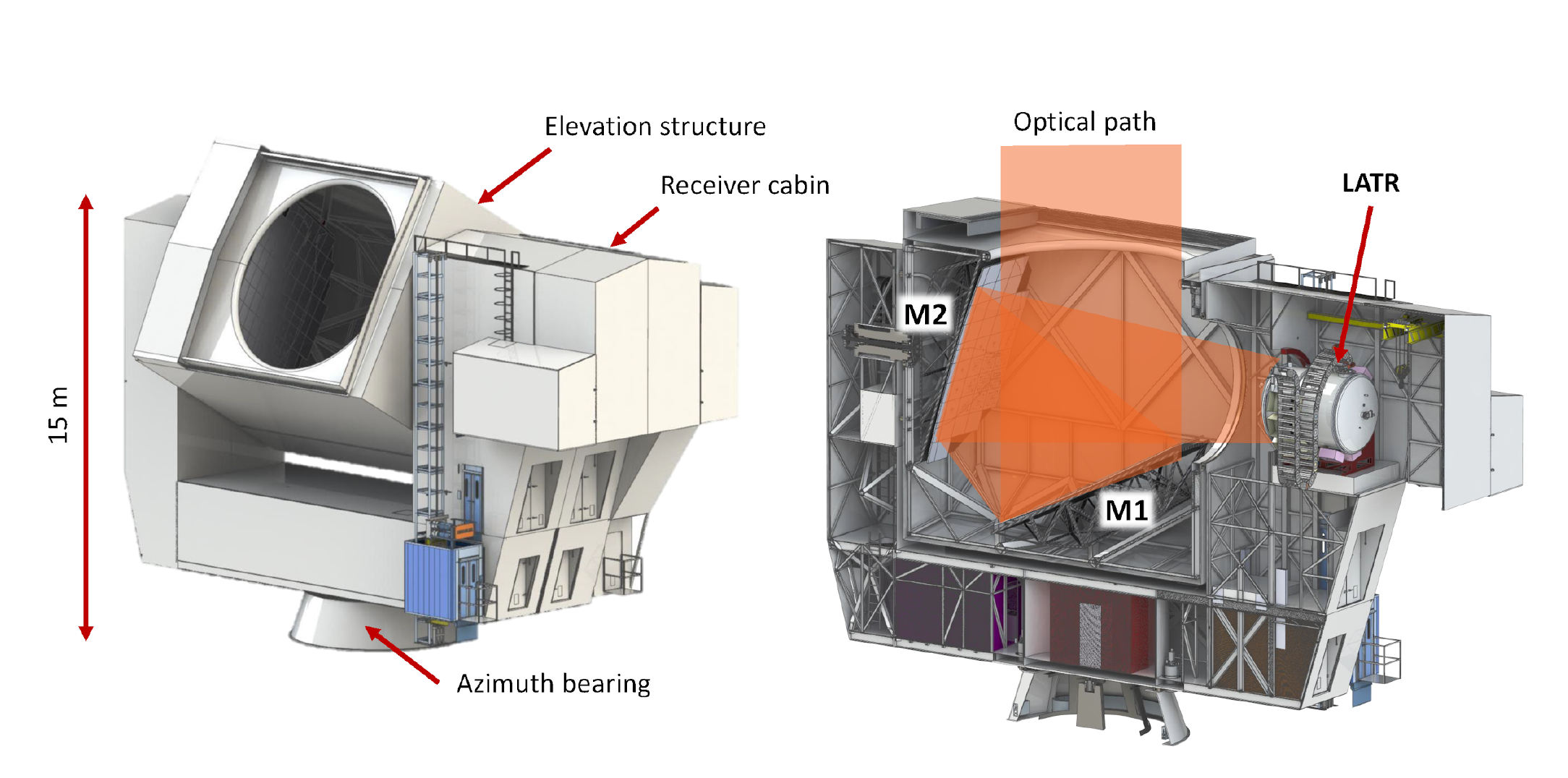}
\caption{Shown is the design for the SO large-aperture telescope (SO LAT).  The Cross-Dragone design of CMB-HD is largely a scaled up version of the SO LAT and CCAT-Prime telescope designs.  The main differences will be the mount for each CMB-HD telescope, since the CMB-HD mounts will need to support more weight, and the addition of a laser metrology system required by CMB-HD. Figure reproduced from~\cite{CMBS4DSR}.}
\label{fig:SO-LAT}  
\end{figure}

The Cross-Dragone design of CMB-HD is largely a scaled up version of the SO large-aperture telescope (LAT) and CCAT-Prime designs, which are currently being manufactured (see Figure~\ref{fig:SO-LAT} for SO LAT design).  The major differences will be the mount for the CMB-HD telescopes, which will need to support more weight than the SO LAT or CCAT-Prime. Another main difference is that CMB-HD will require a laser metrology system to correct for thermal, gravitational, and wind effects on timescales of tens of seconds.  Such a laser metrology system is currently being tested on the GBT 100-meter telescope at millimeter wavelengths.  Table~\ref{tab:telescope} lists the detailed telescope characteristics of CMB-HD.     

\clearpage

\begin{table}[]
\caption{Telescope Characteristics Table for CMB-HD}
\centering
\begin{tabular}{|l|l|}
\hline
{\bf{Telescope}}	    & {\bf{Value/Summary}}	 \\
\hline
\hline
Main and Effective Aperture Size		&  32~meter actual, 28~meter illuminated            \\
\hline
System Effective Focal Length		    &    f/1.9 -- f/2.8                   \\
\hline
Total Collecting Area		            &    500--610 square meters per telescope     \\
&   \\
\hline
Field of View		                    &     $>3.2$ degree diameter at 150~GHz   \\
& (with cold reimaging optics)          \\     
\hline
Wavelength range		                & 0.86 mm to 10 mm          \\
\hline
Optical surface figure quality (RMS)	&  25~$\mu$m requirement (15~$\mu$m goal)~\textsuperscript{a}~~~~~~~~~~                   \\
\hline
Surface Coating Technique		        &   Machined aluminum               \\
\hline
Number of Mirrors or Reflecting Surfaces &    Two                \\	
\hline
Size of each Optical Element 	&    Primary: 32m diameter    \\
and its Clear Aperture & Secondary: 26~m diameter    \\	
\hline
Panel dimensions  &  Primary: 0.75 to 2.0~m  \\
& Secondary: 0.75 to 2.0~m \\
\hline
Panel surface rms (each panel)  & $<10$ microns 
\\
\hline
Mass of each Segment or Element 		&     10 kg/m$^2$ per mirror panel                \\
\hline
Total Moving Mass 	(on elevation bearing)~~~~~~~~~~	                &        700 -- 1500 tons           \\
\hline
Total Moving Mass 	(on azimuth bearing)	                &      2000 -- 7000 tons \\
\hline
Mass of each Optical Element 		    &      7000 kg mirror alone, \\ 
& 14 tons with support structure~\textsuperscript{b}~~ 
            \\
\hline
Panel actuator Precision and Range		    &     Course manual adjust +-25cm; \\
& fine (5$\mu$m precision) adjust +-1~cm~\textsuperscript{c}~~~~~~~~~~~~               \\
\hline
\end{tabular}
\begin{tablenotes}[flushleft]
\item \textsuperscript{a} Commercial laser trackers can achieve better than 10~$\mu$m measurements along a line of sight over 30 meters; a similar system is being deployed at the GBT.  We would expect such a system to correct errors in real time at a panel level taking care of thermal, gravitational, and wind effects on timescales of tens of seconds.   

\item \textsuperscript{b} 
We baseline a carbon backup structure, and can descope to the same material as the panels if thermal and cost constraints allow; a full study to minimize panel gaps while ensuring that they never drop to zero is needed and will affect the material choice. 

\item \textsuperscript{c}
The range needs to be enough to take into account all thermal and gravitational distortions with an accuracy better than our measurement of the panel location.  Lighter (more flexible but cheaper) telescope structures are possible if real-time adjustment/measurement of mirror shapes and locations are better.   
\end{tablenotes}
\label{tab:telescope}
\end{table}
\clearpage
\begin{table}[h]
\setcounter{table}{3}
\caption{Telescope Characteristics Table for CMB-HD (continued)}
\centering
\begin{tabular}{|l|l|}
\hline
{\bf{Telescope}}	    & {\bf{Value/Summary}}	 \\
\hline
\hline
Degrees of Freedom (mirror panels)		                & 5 actuators (corners/center) in Z               \\
& and manual x,y \\
\hline
Degrees of Freedom (mirror) & 5  (x,y,z,tiptilt,wedge)~\textsuperscript{d}~~ \\ \hline
Type of Mount used for Pointing and Allowed Range~~~~~ &  az-el, -40$<$az$<$400; 0$<$el$<$180                   \\		
\hline
Mass and Type of Material for Support Structure &   7 tons of carbon fiber (each mirror)~~~ \\
& steel telescope structure~\textsuperscript{e}~~  \\ 		
\hline
Optic Design 		    &   Cross-Dragone                  \\
\hline
Description of Adaptive Optics	    &  Laser metrology system   \\ 
\hline
\end{tabular}
\begin{tablenotes}[flushleft]
\item \textsuperscript{d}
Note that to keep the mirrors fixed with respect to each other the secondary mirror may be moved to obtain and maintain alignment.
\item \textsuperscript{e}
With an active measurement system one does not need low CTE materials; steel is much more cost effective.
\end{tablenotes}
\end{table}

\vspace{-2mm}
\subsubsection{Receiver}

The CMB-HD telescope cameras will hold about 400,000 pixels.  Each pixel will have two frequency bands and two polarizations for a total of 1.6~million detectors. We assume in the baseline design horn-fed TES detectors, however, MKIDs may also be a viable detector technology, which could reduce the cost significantly. MKIDs are less complex to fabricate and are naturally multiplexed, simplifying their readout. We note that MKIDs are currently being used by MUSTANG-2 at 90 GHz, and TolTEC will use MKIDs at 150, 220, and 270 GHz; thus they are currently being field tested at similar frequencies and resolution as CMB-HD.  Using MKIDs at lower frequencies would require further MKID development. CMB-HD will have four frequency band pairs:~30/40, 90/150, 220/280, 280/350 GHz.  The distribution of detectors per frequency for the first three band pairs will be similar to the ratios adopted by SO, and achieve the noise levels given above.  These ratios were calculated by calculating target map noise for each frequency from the science requirements and then taking into account how detector noise varies with frequency. \\

\begin{figure}[t]
\centering
\includegraphics[width=0.5\textwidth]{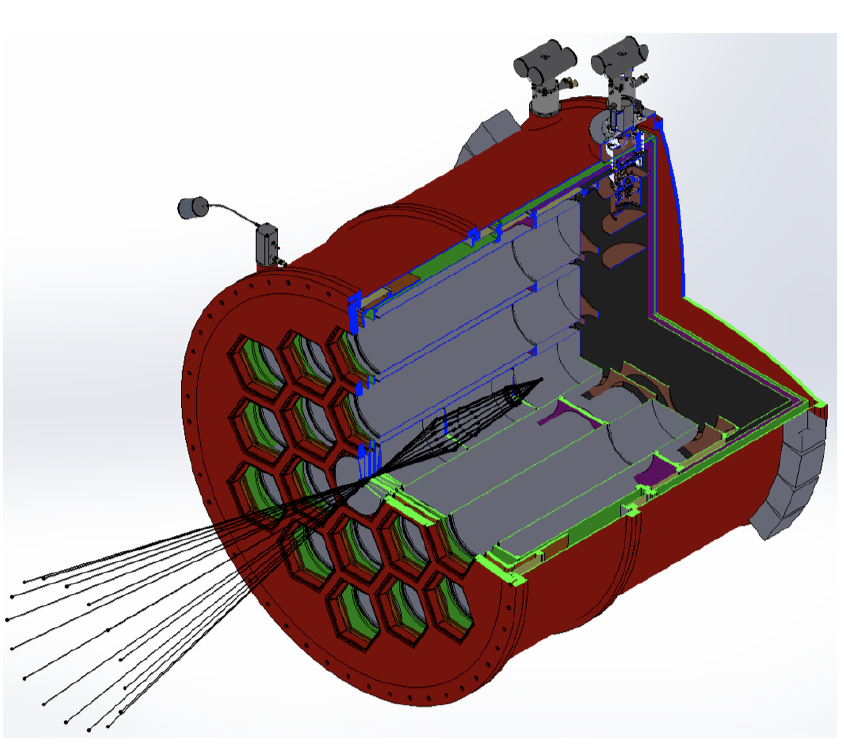}
\caption{CMB-HD will adopt a similar design for its baseline receiver as that used by SO and CCAT-Prime. Shown here is the CMB-S4 receiver design, which is an extension of the SO LATR design; the CMB-S4 receiver has a 2.6 meter diameter and 19 optics tubes, whereas the SO LATR has 13 optics tubes.  With the aid of a warm first lens, it is possible to focus light through smaller vacuum windows with very little gaps between tubes - currently tens of centimeters are needed between tubes in order to maintain the strength of the cryostat resulting in a loss of ~50\% of the telescope's focal plane. CMB-HD will aim to increase the packing density of optics tubes. Seven receivers are assumed to fit in the CMB-HD focal plane, which has a diameter of about 7.3 meters, for each telescope. Figure reproduced from~\cite{CMBS4DSR}.}
\label{fig:receiver}  
\end{figure}

For a baseline receiver we adopt a design similar to that for CCAT-prime and SO (see Figure~\ref{fig:receiver}).  Multiple sets of cold silicon lenses re-image the telescope focal plane while adding a 1K lyot stop, baffles for control of stray light, and cold blocking and bandpass filters.  For SO and CCAT-prime these are housed in a single cryostat. However, for a telescope the size CMB-HD it makes sense to group them into a number of cryostats; a single cryostat many meters across is hard to build and impractical to transport and maintain. In addition, recent advances in low-loss silicon~\cite{Chesmore2018,Wollack2020} could allow a warm first lens, and in that case the packing density of tubes could become far greater~\cite{Niemack2016}, allowing for smaller cheaper cryostats. \\

The baseline design of CMB-HD assumes seven cryostats for each telescope, each of which has a diameter of about 2.5 meters (including flanges).  The focal plane area is about 7.3 meters in diameter, which can contain the seven cryostats.  Each cryostat holds between 13 and 26 optics tubes (depending on a final optimization of tube size versus cost and focal plane usage).  Every optics tube has a vacuum window, thermal-infrared blocking filters, and cold lenses that focus the light onto an array of detectors cooled to 0.1K.  Each optics tube will hold 500 to 4000 pixels (depending on frequency) that are sensitive to two polarizations and two frequency bands per pixel (i.e.~four detectors per pixel).  Assuming $2f*\lambda$ spacing of detectors and a conservative filling factor of the focal plane of 50\% (due to gaps between detector wafers, optics tubes, and cryostats), then with the optical throughput of our design it is possible to fit $\sim 132,000$ pixels per telescope. We will aim for a more optimistic filling factor of 75\%, yielding about 200,000 pixels per telescope.\\ 

Other than achieving the high packing density of detectors in the receivers, the assumed instrumentation is based on demonstrated technology.  Similar optics tubes have been used by ACT, BICEP/Keck, POLARBEAR, and SPT.  ACT has demonstrated successful use of horn-coupled dichroic, dual-polarization pixels.  ACT has also successfully cooled their detector arrays to 0.1K using a dilution refrigerator. Pulse-tube cryocoolers cool the optics and thermal shields to 4K.  Horn-coupled TES detectors read out by time-domain multiplexing have the highest technology readiness level and have been used in CMB experiments like ABS, ACTPol, Advanced ACTPol (AdvACT), CLASS, and SPTpol.  CMB-HD will cover a spectral range of 30 to 350 GHz.  Table~\ref{tab:inst} lists the characteristics of the receiver instrumentation.

\begin{table}[t]
\caption{{} Receiver Instrumentation for CMB-HD{}}
\centering
\begin{tabular}{|l|l|}
\hline
{\bf{Item}}                                         & {\bf{Value}}      \\
\hline
\hline
Type of Instrument                           &    Polarization-sensitive bolometer cameras;            \\
& Seven cameras per telescope \\
\hline
Spectral Range                               &       Dichroic pixels at 30/40, 90/150,             \\
&  220/280, and 280/350 GHz \\
\hline
Optics Tubes &  91 to 182 per telescope,  \\
& each tube with about 40 cm clear aperture; \\
& Distributed roughly in the ratio of 1:4:2:2 for\\
& 30/40, 90/150, 220/280, and 280/350 respectively \\ 
\hline
Number of Detectors, type, and pixel count    &    1.6 million TES bolometers (400,000 pixels)              \\
\hline
Thermal or Cryogenic Requirements            &  0.1K (Pulse tubes to 4K + Dilution fridge)                \\
\hline
Size/Dimensions   (for each instrument)                           &   2.5m diameter,  2--3m long      \\
\hline
Instrument average science data volume / day &   190 TB/day uncompressed               \\
\hline
Instrument Field of View                     &   1 degree diameter (each)               \\
\hline
Development Schedule                         &   2 years design + 2 years  construction \\
& (construction schedule will be limited by \\
& detector/readout fabrication)        \\
\hline
\end{tabular}
\label{tab:inst}
\end{table}

\subsubsection{Site}

\begin{figure*}[t]
  \centering
  \includegraphics[width=0.80\textwidth]{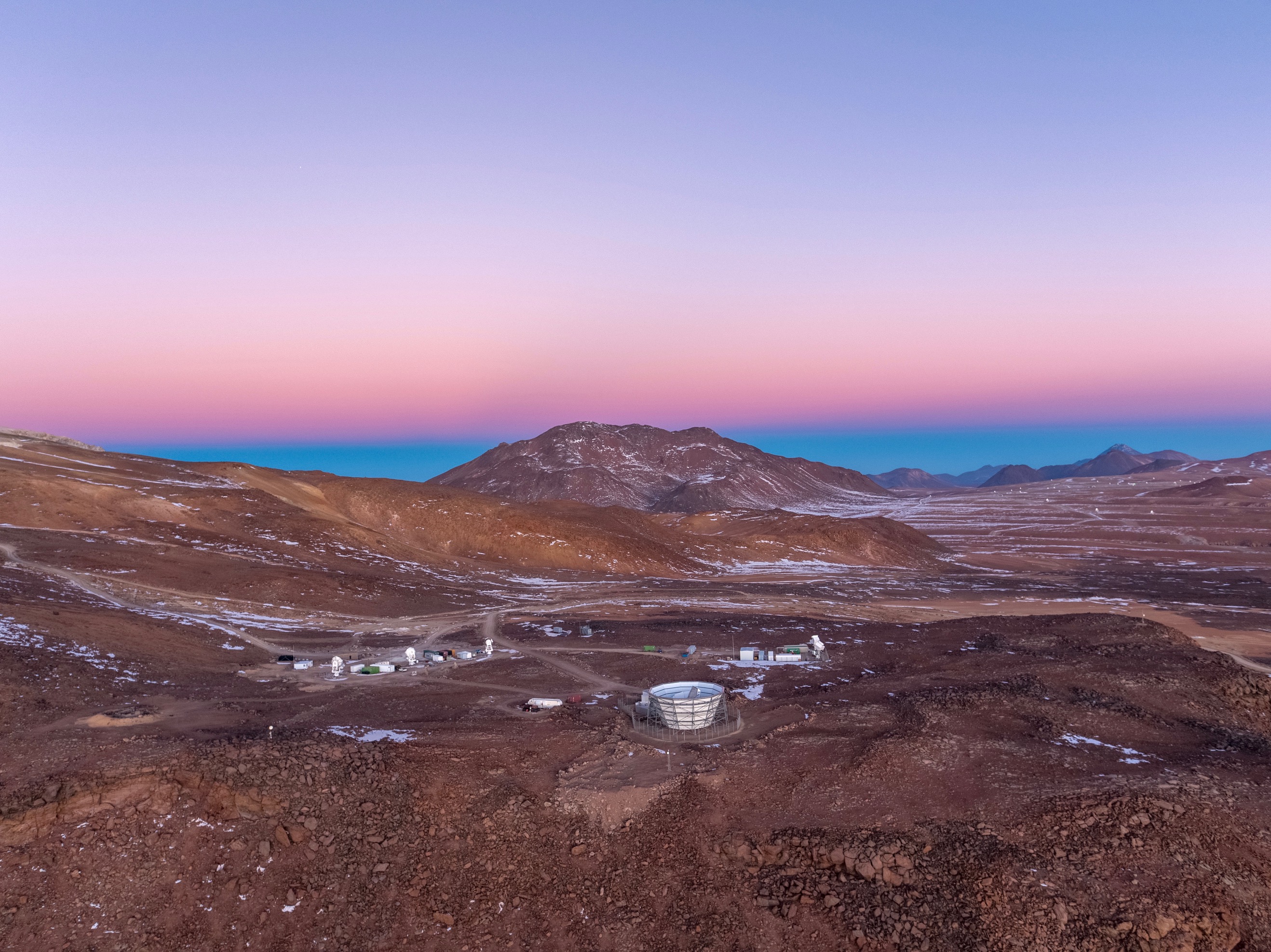}
  \includegraphics[width=0.80\textwidth]{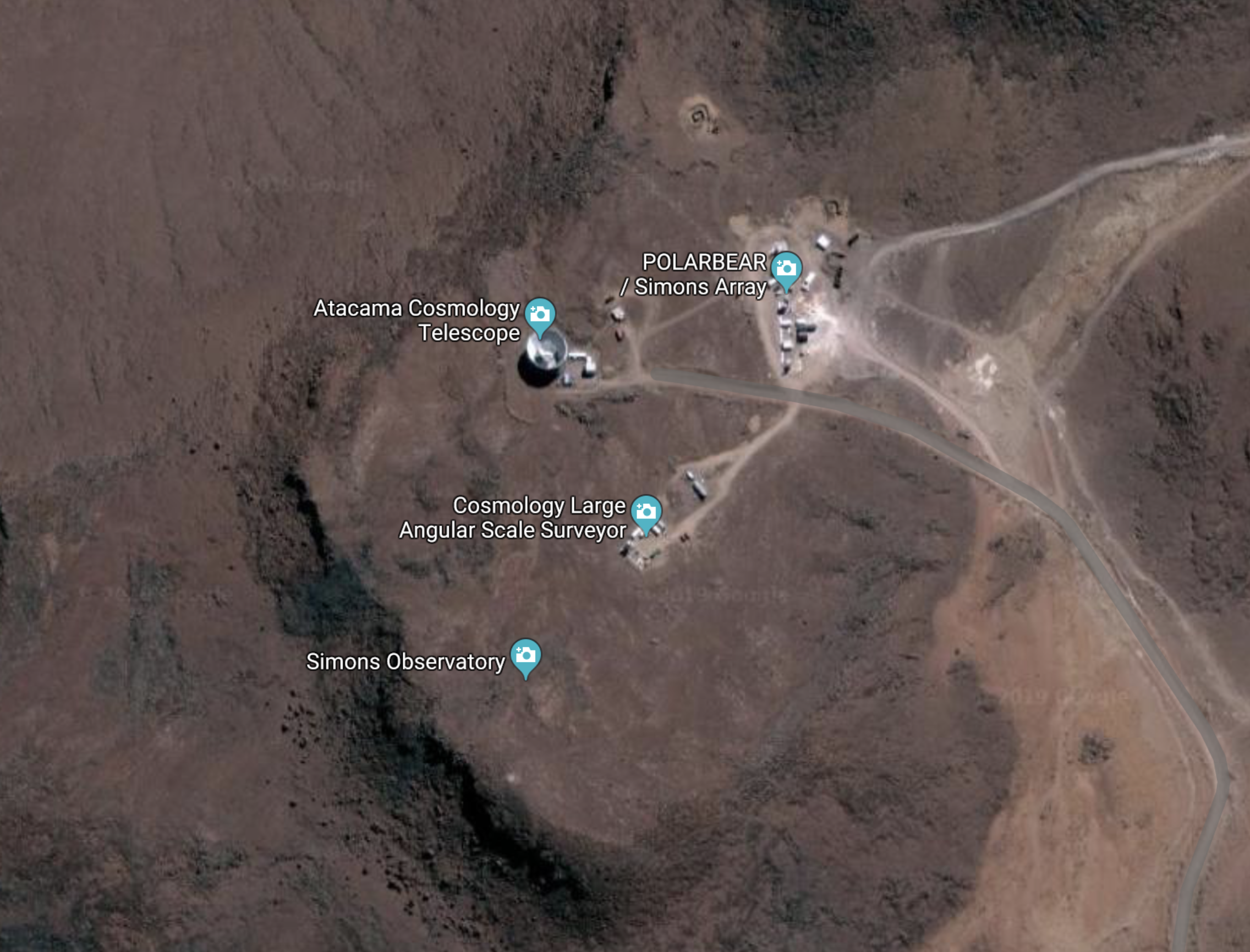}
  \caption{Cerro Toco in the Atacama Desert. {\it{Top:}} Photo courtesy of {\it{Debra Kellner}}. {\it{Bottom:}} Image from Google Earth.  Telescopes pictured are the Atacama Cosmology Telescope (ACT), POLARBEAR/Simons Array, and the Cosmology Large Angular Scale Surveyor (CLASS). Also shown is the site of the funded Simons Observatory (SO).}
  \label{fig:site}
\end{figure*}

Cerro Toco in the Atacama Desert is the best site to achieve the science of CMB-HD \citep[e.g.][]{Gomez-Toribio2022}.  It is 5200 meters above sea level and sits on public Atacama Astronomical Park (AAP) land.  An instrument at this site can survey the required 50\% of the sky.  The sensitivity requirement of 0.5~$\mu$K-arcmin also requires locating CMB-HD at a high, dry site with low precipitable water vapor to minimize the total number of detectors needed.  No site within the U.S. has a suitable atmosphere.  While a higher site than Cerro Toco, such as Cerro Chajnantor in the Atacama Desert, might reduce the detector count further, that likely does not outweigh the increased cost of construction and operations at the higher site.  \\

Figure~\ref{fig:site} shows the existing CMB telescopes already on this plateau, namely ACT, POLARBEAR/Simons Array, and the Cosmology Large Angular Scale Surveyor (CLASS). Also shown is the location of the funded SO, which is expected to have first light in 2024.  In addition to the footprints of these existing and planned telescope facilities, there is about 300,000 square feet of available level ground at the Cerro Toco site that is appropriate for supporting telescopes and accompanying buildings.  SO will build two new buildings at this site, one that includes a high-bay lab (11m x 8m x 8.5m high) that can support an 8-ton crane, as well as one that includes a control room and office. These buildings are designed to satisfy engineering code IBC-2009, with maximum allowed winds of 90 mph. CMB-HD would build at least two new buildings, roughly similar in scope to the SO buildings.  \\

Regarding the atmospheric characteristics of the site, Cerro Toco is one of the driest places on Earth, which is a necessity to achieve the science of CMB-HD.  This site has high atmospheric transmittance and low atmospheric emission across the relevant millimeter-wave frequencies; this has been studied extensively as discussed in~\cite{Bustos2014, Lay2000}.  The Cerro Toco site has been used for over a decade by ACT and POLARBEAR/Simons Array.  As a result, how the atmosphere impacts the noise properties of existing instruments has been well quantified, and this information was propagated into the detector requirements for CMB-HD discussed in Section~\ref{sec:instReq}. \\

\begin{table}[t]
\caption{{} CMB-HD WBS at Level 2 with Definitions {}}
\centering
\begin{tabular}{|l|l|}
\hline
{\bf{Work Breakdown Structure}}                                         & {\bf{Definitions}}      \\
\hline
\hline
1.1 - Project Management       &      Management and systems engineering during the \\
& construction phase        \\
\hline
1.2 - Telescopes    &  Materials, equipment, labor, and travel associated with \\
& the design and construction of the telescopes   \\
\hline
1.3 - Telescope Receivers &  Materials, equipment, labor, and travel associated with \\
& the design, fabrication, assembly, and testing of the receivers\\
\hline
1.4 - Detectors and Readout &    Fabrication, assembly, and testing of the detectors and \\
& cold and warm readout electronics          \\
\hline
1.5 - Data Acquisition            &    Delivery of the control systems for the observatories \\
& and data acquisition    \\
\hline
1.6 - Data Management                         &   Maintenance of site computing,
networking and \\  
& data storage; staff for data acquisition,
pipeline \\
& development, and map making   \\
\hline
1.7 - Site Infrastructure &   Materials, equipment, labor, and travel needed to manage \\
& and oversee construction activities at the site             \\
\hline
1.8 - Integration and Testing    & On-site integration and commissioning of the telescopes\\
& and infrastructure         \\
\hline    

\hline
\end{tabular}
\label{tab:wbs}
\end{table}

\begin{table}[t]
\caption{{} Project Cost Estimates for CMB-HD Instrumentation and Construction — US Only {}}
\centering
\begin{tabular}{|l|c|c|c|c|c|c|c|c|c|c|c|}
\hline
Project Item (WBS 1)                                        & Y1 & Y2 & Y3 & Y4 & Y5 & Y6 & Y7 & Y8 & Y9 & Y10 & Total (\$M)        \\
\hline
\hline
1.1 Project Management     & 10  & 10  & 10 & 10 & 10 & 10 & 10  & 10  & 10 & 10 & 100  \\
\hline
1.2 Telescopes       &   &   &  &  &  & 400 &   &   &  &  &      400   \\
\hline
1.3 Telescope Receivers &   &   &  & 20 & 20 & 20 & 20  & 20   &  &  &      100   \\
\hline
1.4 Detectors \& Readout    &   &   &   & 64  & 64 & 64 & 64  & 64  &  &  &  320  \\
\hline
1.5 Data Acquisition   &   &   & 1 & 1 & 1 &1 & 1  &  1 & 1 & 1 &  8 \\
\hline
1.6 Data Management  & 2.4  & 2.4  & 2.4 & 2.4 & 2.4 &2.4 & 2.4  & 2.4  & 2.4 & 2.4 &   24  \\
\hline
1.7 Site Infrastructure &   &   &  & 10  & 10 & 6 & 6  & 6  & 6 & 6 &    50      \\
\hline
1.8 Integration \& Testing  &   &   &  &  & 4 & 4 & 4  & 4  & 4 &  &   20  \\
\hline
Total  &   &   &  &  &  & &   &   &  &  &   1022      \\
\hline
\end{tabular}
\label{tab:cost}
\end{table}

\subsection{Cost Estimates}

Table~\ref{tab:wbs} provides the CMB-HD WBS at Level 2 with definitions.  The CMB-HD instrument and site cost is shown in Table~\ref{tab:cost} with 2019 estimates. Much of the cost is in fabricating the detectors and their readout electronics.  For detector cost comparison, CMB-HD will have a factor of 3.2 more detectors than the proposed CMB-S4 experiment (1600k for \$320 million versus 500k for \$100 million).  However, there are two avenues that could potentially reduce this cost.  One is that over the next several years, as upcoming experiments such as SO require tens of thousands of detectors, the mass production of detectors may drop the cost.  The other is that MKIDs, a detector technology alternate to TES devices, are currently being tested on-sky at millimeter wavelengths~\cite{Austermann2018}. If MKID detector technology matches in practice to its promise, then the detector cost may drop by a factor of a few due to simplified readout and fabrication. Thus, further development of MKID technologies and high frequency (350 GHz) technologies in the coming years will further enable this project.\\

Two 30-meter-scale off-axis, crossed Dragone telescopes are costed at \$400 million. The telescope receivers are estimated at \$100 million (roughly a factor of 3 higher than estimated for CMB-S4).  Project management is estimated at \$10 million per year over ten years, and site infrastructure is estimated at \$50 million total. Data management and data acquisition are estimated at 12 FTEs/year and 4 FTEs/year, respectively, assuming \$200,000 per FTE (including 40\% for benefits and 60\% for overheads).

\section{Summary}

CMB-HD is an ambitious leap beyond previous and upcoming ground-based millimeter-wave experiments. It will allow us to cross critical measurement thresholds and definitively answer pressing questions about dark matter and dark sectors, light particle species, inflation, dark energy, neutrino mass, and other beyond standard model physics.  In addition, CMB-HD can open new windows on galaxy evolution, planetary studies, and transient phenomena.  The CMB-HD project also recognizes the productivity benefit of an engaged collaboration, and is committed to fostering a culture that enables this.  The CMB-HD survey will be made publicly available, and the project will prioritize usability and accessibility of the data by the broader scientific community.

\clearpage
\addcontentsline{toc}{section}{References}
\bibliographystyle{unsrturltrunc6}

\bibliography{cmb-in-hd.bib}

\end{document}